\documentclass[iop,numberedappendix,twocolappendix]{emulateapj}
\usepackage{natbib}
\usepackage[pass,letterpaper]{geometry}

\newcommand{\set}[1]{\ensuremath{\left\{\,#1\,\right\}}}

\newcommand{\msun}{$M_{\odot}$}
\newcommand{\midline}{\, | \,}

\slugcomment{Accepted for publication in the ApJ.}
\begin{document}

\shortauthors{Gordon et al.}
\shorttitle{The BEAST}

\title{The Panchromatic Hubble Andromeda Treasury XV. \\
The BEAST: Bayesian Extinction and Stellar Tool
\footnote{Based on observations made with the NASA/ESA Hubble
Space Telescope, obtained from the Data Archive at the Space Telescope
Science Institute, which is operated by the Association of
Universities for Research in Astronomy, Inc., under NASA contract NAS
5-26555.}}

\author{Karl~D.~Gordon\altaffilmark{1,2}, 
   Morgan~Fouesneau\altaffilmark{3}, 
   Heddy~Arab\altaffilmark{1},
   Kirill~Tchernyshyov\altaffilmark{4},
   Daniel~R.~Weisz\altaffilmark{6,7}, 
   Julianne~J.~Dalcanton\altaffilmark{6},
   Benjamin~F.~Williams\altaffilmark{6},
   Eric~F.~Bell\altaffilmark{14},
   Luciana~Bianchi\altaffilmark{4},
   Martha~Boyer\altaffilmark{11,12},
   Yumi~Choi\altaffilmark{6},
   Andrew~Dolphin\altaffilmark{8},
   L\'eo~Girardi\altaffilmark{10}, 
   David~W.~Hogg\altaffilmark{9,3}, 
   Jason~S.~Kalirai\altaffilmark{1},
   Maria~Kapala\altaffilmark{3},
   Alexia~R.~Lewis\altaffilmark{6},
   Hans-Walter~Rix\altaffilmark{3},
   Karin~Sandstrom\altaffilmark{13},
   \&
   Evan~D.~Skillman\altaffilmark{15}}

\altaffiltext{1}{Space Telescope Science Institute, 3700 San Martin
  Drive, Baltimore, MD 21218, USA; kgordon@stsci.edu} 
\altaffiltext{2}{Sterrenkundig Observatorium, Universiteit Gent, 
  Krijgslaan 281 S9, B-9000 Gent, Belgium}
\altaffiltext{3}{Max Planck Institute for Astronomy, Koenigstuhl 17, 
  69117 Heidelberg, Germany}
\altaffiltext{4}{Department of Physics and Astronomy, Johns Hopkins
  University, Baltimore, MD 21218, USA}
\altaffiltext{6}{Department of Astronomy, University of Washington, 
  Box 351580, Seattle, WA 98195, USA}
\altaffiltext{7}{Hubble Fellow}
\altaffiltext{8}{Raytheon Company, Tucson, AX, 85734, USA}
\altaffiltext{9}{Center for Cosmology and Particle Physics, New York 
  University, 4 Washington Place, New York, NY 10003, USA}
\altaffiltext{10}{Osservatorio Astronomico di Padova-INAF, Vicolo dell'Osservatorio 5, 
  I-35122 Padova, Italy}
\altaffiltext{11}{Observational Cosmology Lab, Code 665, NASA Goddard
  Space Flight Center, Greenbelt, MD 20771, USA}
\altaffiltext{12}{Oak Ridge Associated Universities (ORAU), Oak Ridge,
  TN 37831, USA}
\altaffiltext{13}{Steward Observatory, University of Arizona, 933
  North Cherry Avenue, Tucson, AZ, 85721, USA}
\altaffiltext{14}{Department of Astronomy, University of Michigan, 500
  Church Street, Ann Arbor, MI 48109, USA}
\altaffiltext{15}{Minnesota Institute for Astrophysics, University of
  Minnesota, Minneapolis, MN 55455, USA}

\begin{abstract} 
We present the Bayesian Extinction And Stellar Tool (BEAST), a
probabilistic approach to modeling the dust extinguished photometric
spectral energy distribution of an individual star while accounting
for observational uncertainties common to large resolved star surveys.
Given a set of photometric measurements and an observational uncertainty
model, the BEAST infers the physical properties of the stellar source
using stellar evolution and atmosphere models and constrains the line
of sight extinction using a newly developed mixture model that
encompasses the full range of dust extinction curves seen in the Local
Group.  The BEAST is specifically formulated for use with large
multi-band surveys of resolved stellar populations.  Our
approach accounts for measurement uncertainties and any
covariance between them due to stellar crowding (both systematic
biases and uncertainties in the bias) and absolute flux calibration,
thereby incorporating the full information content of the measurement.
We illustrate the accuracy and precision possible with the BEAST using
data from the Panchromatic Hubble Andromeda Treasury. While the BEAST 
has been developed for this survey, it can be easily applied to
similar existing and planned resolved star surveys.
\end{abstract}


\section{Introduction}
\label{sec_intro}

The ability to resolve the stellar content of numerous galaxies has
revolutionized our understanding of galaxy formation and evolution.
The color and luminosity of an individual star encodes information
about the star's intrinsic mass, age, and metallicity and further
illuminates the intervening dust, revealing the composition, grain
size distribution, and column density of the obscuring material.
Decoding this information provides new constraints on the mass
assembly and chemical history of a galaxy and on the detailed study
of a galaxy's interstellar medium structure and dust composition.  The
high information content contained in resolved stars has motivated
extensive surveys that have cataloged hundreds of millions of stars in
the Milky Way and Local Volume \citep[e.g.,][]{zaritsky1997,
  stoughton2002, holtzman2006, mcconnachie2009, dalcanton2009,
  dalcanton2012a, sabbi13, Williams14}.  This number will increase at
least ten-fold with the next generation of dedicated programs (e.g.,
LSST, Gaia, PAN-STARRS, etc.).

\begin{figure*}[tbp] 
\epsscale{1.15}
\plotone{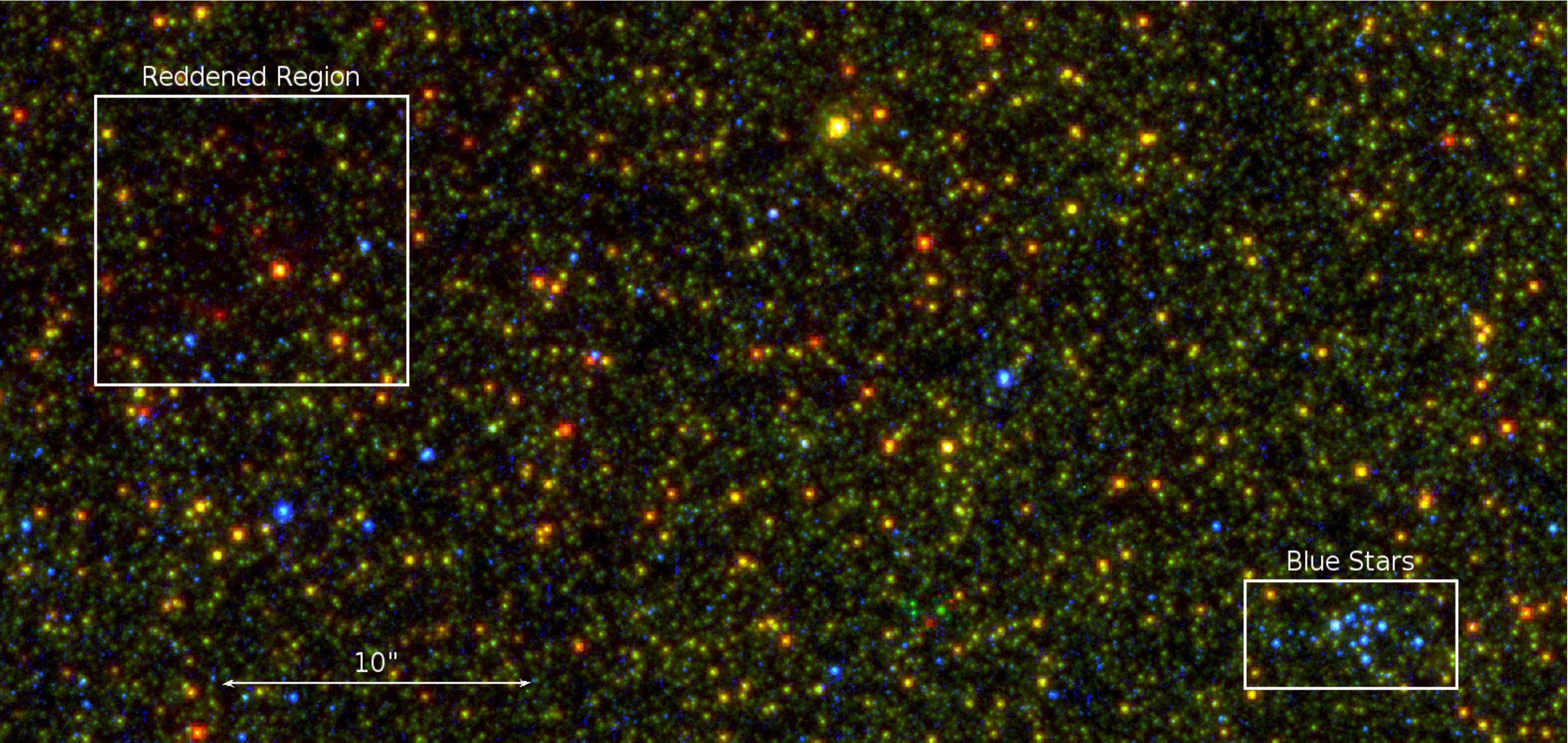}
\caption{A small region from the PHAT survey of M31 in the 10~kpc star
  forming ring is shown using the F336W (blue), F814W (green), and
  F160W (red) band images.  The entire region shows strong crowding
  with sources very near each other.  In addition, the diverse range
  of environments is illustrated by the two marked regions, one
  showing a region that suffers from dust reddening (red sources and
  fewer sources overall) and the other a blue cluster of stars.
\label{fig_3col_image_ex} }
\end{figure*}

With the wealth of resolved star data in external galaxies, both
available and forthcoming, 
it is essential to develop robust methods that can fully exploit observables
to infer well-characterized physical quantities. Here we concentrate
on the observed spectral 
energy distributions (SED) of single stars, which reflects three different
aspects: stellar physics, intervening dust, and observational effects.
Individual stars emit photons over a wide range of wavelengths.  A
portion of these photons are then removed by dust through absorption
and scattering out of the line-of-sight, extinguishing the star's
intrinsic spectrum.  Finally, the precision and accuracy of the
measured flux is modulated by the finite number of observed
photons, the contrast between the target and local background, the
wavelength sensitivity of the selected instrument (e.g., filter
sensitivity in the case of photometric observations), and the accuracy
with which the flux from neighboring stars can be subtracted.
Therefore, recovering the intrinsic properties of the star and the
intervening dust requires modeling of both physical and observational
effects.

Historically, stellar SED fitting techniques have taken simplified
approaches that make this analysis more tractable, especially for
observations of sources in our Galaxy.  These compromises
often involve focusing on singular science goals such as constraining
only stellar metallicities, e.g., the `Ultraviolet excess technique'
\citep[e.g.,][]{wallerstein1960}, or removing line of sight extinction
effects, e.g., the `Q' parameter \citep[e.g.,][]{johnson1953}.
Additionally, the characterization of uncertainties is often limited
only to photon noise and absolute flux calibration uncertainties, 
motivating the use of conventional $\chi^2$ fitting techniques.  
However these
approximations can strongly compromise the results
from resolved surveys of external galaxies where they are not valid.

For the most part, such techniques have provided insightful
astrophysical results using various Galactic and extragalactic
datasets \citep{bianchi2001, romaniello2002, zaritsky2004,
  berry2012, schlafly11, bianchi12b}.  However, these methods have a
number of shortcomings that hinder their ability to completely exploit
the information content of observations, including the inability to
accurately model data in moderate and low signal-to-noise regimes or
to incorporate full accounting of the observational uncertainties.
Thus, in the era of large, deep, and expensive surveys, we should not limit
ourselves to sub-optimal fitting methods to analyze resulting
datasets.
 
In the past decade, the introduction of probabilistic stellar SED
fitting techniques has led to significant improvements
\citep{maizappellaniz2004, Fouesneau10, Bailer-Jones11,
  bianchi12a, bianchi12b, dario2012, green14, Schonrich14, Ness15}.
These approaches are 
designed to better capture crucial inter-parameter degeneracies and
provide flexible frameworks for including different models of
stellar evolution, stellar atmospheres, and extinction laws.  However,
despite the marked improvement in techniques, the current generation
of stellar SED fitting codes are still not optimal for analyzing
datasets generated by the current and next-generation resolved star
surveys of external galaxies.

One limitation of existing SED fitting codes is in their treatment of
dust extinction.  Current and new datasets cover diverse ranges in
galactic environments showing a range of dust content, necessitating a
comprehensive model of interstellar dust extinction to correctly
constrain both dust and stellar parameters.  A second limitation of
current SED fitting codes is in their treatment of source ``crowding''
wherein the measured flux of a source is affected by the presence
of nearby sources.  As a result, recovered stellar fluxes are affected
by systematic biases and uncertainties due to local crowding that are
beyond those expected for random photometric errors due to photon
counts \citep{stetson88}.  Such systematics can and usually do dominate
the photometric error budget for most stars in extragalactic surveys.  These
effects can be particularly large in some of the most scientifically
interesting regions such as stellar clusters and star forming
complexes.

Large datasets of resolved stars in external galaxies contain a
wealth of information near a given survey's 
detection limit, where crowding often induces strong correlations in
the measurement uncertainties and becomes the dominant source of
uncertainty.  Fig.~\ref{fig_3col_image_ex} gives an example image from
the Panchromatic Hubble Andromeda Treasury
\citep[PHAT;][]{dalcanton2012b} program that shows strong crowding and
a range of dust extinctions.  Failure to account for these effects can
lead to systematically incorrect inferences about parameters and an
inability to fully capture the complete information content of the
observations, undermining our ability to correctly interpret large
surveys.

Finally, survey observations cover a wide range in stellar spectral
types and dust extinctions, implying that not all stars will be
detected at a high degree of significance in all filters.  For
example, the ultraviolet (UV) flux of luminous cool asymptotic giant
branch stars can easily be fainter than the observational limit of
the UV bands in a survey.  On the other hand, a $\sim$5~\msun\ main sequence star
may be recovered with moderate signal-to-noise in the UV, but have low
signal-to-noise ($<$3$\sigma$) in the near-infrared.  Moreover even
low signal-to-noise detections have important information about the SED and
therefore should {\bf be} included when constraining the stellar and dust
parameters of a survey.

Our goal is to devise an approach for recovering the intrinsic physical
properties of a star (e.g., mass, age, metallicity, etc.) and the
intervening dust (e.g., composition, column density, and size
distribution) while simultaneously including robust uncertainties on
each parameter given the observed SED and known observational effects.
To accomplish this goal, we present the probabilistic framework for
the Bayesian Extinction and Stellar Tool (BEAST).  The development of
this methodology has been motivated by the PHAT program
\citep{dalcanton2012b}, an 828-orbit HST multi-cycle program that has
observed $\sim$1/3 of Andromeda's star-forming disk from the UV
through the near-infrared (NIR).  PHAT has cataloged 6-band fluxes for
$>$100 million individual stars \citep{Williams14}, forming a critical
dataset for better understanding the relationship between star
formation and the evolution of M31's baryonic content.  The guiding
principle of the BEAST is to accurately derive the stellar and dust
parameters in all signal-to-noise regimes, and report well
characterized uncertainties.  Thus results from the BEAST allow for the study
of individually well measured stars as well as statistical studies
that take advantage of the large number of sources detected in modern
surveys.

We start in \S\ref{sec:likelihood} with the details of our fitting
technique, explicitly including correlations in the uncertainties
between the observed bands using a multivariate Gaussian
distribution.  In \S\ref{sec_sed_model}, we provide the details of our
model for stellar SEDs and develop a mixture model for the dust
extinction.  As an example of applying the BEAST to a specific
dataset, in \S\ref{sec_phat_details} we give the implementation
details for the PHAT data.  Examples of BEAST results using
PHAT data are given in \S\ref{sec_phat_results}.

\begin{figure}[tbp] 
\epsscale{1.2}
\plotone{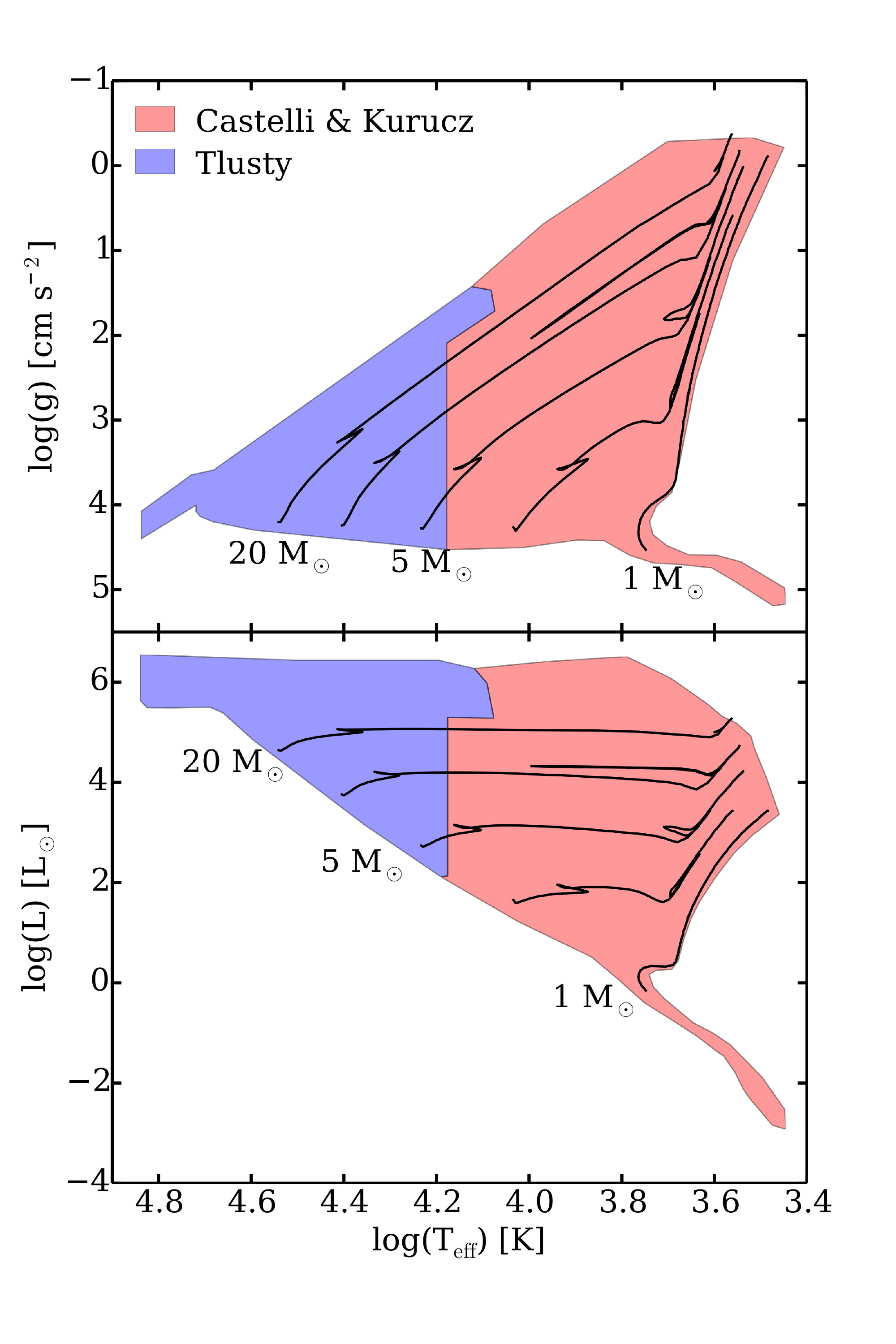}
\caption{The coverage of the stellar models used in the BEAST are
  shown in a Hertzsprung-Russell diagram (top) and
  $\log(T_\mathrm{eff})$--$\log(g)$ plot (bottom).  The two stellar
  atmosphere models used are \citet{Castelli04} and TLusty
  \citep{Lanz03, Lanz07}.  Padova stellar evolutionary tracks
  \citep{Marigo08, Girardi10} for representative stellar masses are
  shown as solid black lines for reference.
\label{fig_stellar_mod_overlap} }
\end{figure}

\section{Fitting Technique}
\label{sec:likelihood}

We undertake a probabilistic approach to modeling the SED of a single
star.  Probability theory provides an established framework for
merging the information of multiple separate models (e.g., physical
models of stars and dust), comparing the resulting model with
observations, and tracking all sources of uncertainty, including
covariances between parameters.

Our data consist of $N$ photometric measurements of a single source.
We designate this set of flux measurements as $F_D$, and then write
the probability of observing $F_D$ given our model parameters,
$\theta$, using a multivariate Normal/Gaussian distribution as
\begin{equation}
  P(F_D \midline \theta) \,=\,
  \frac{1}{Q(\theta)}\,\exp\left(-\frac{1}{2}\chi^2(\theta)\right) \, , \\
  \label{eq:likelihood}
\end{equation}
where 
\begin{equation}
  Q^2(\theta) \,\equiv\, (2\pi)^N \mathrm{det}\left| \mathbb{C}(\theta) \right| \, ,
\end{equation}
\begin{equation}
  \chi^2(\theta) \,=\, \Delta^T \mathbb{C}(\theta)^{-1} \Delta \, , 
  \label{eq:chi2_error}
\end{equation}
\begin{equation}
 \Delta = F_D - F_M(\theta) + \mu(\theta) \, ,
\end{equation}
$\mathbb{C}(\theta)$ is the covariance matrix of the $N$ photometric
bands\footnote{We have chosen to use the notation $\mathbb{C}(\theta)$
for the covariance matrix instead of the more standard $\Sigma$ to
avoid confusion with the use of $\Sigma$ as the summation symbol.},
$F_M(\theta)$ is the predicted flux in the $N$ photometric bands, and
$\mu(\theta)$ is the crowding bias in each band.  The use of a
multivariate Normal/Gaussian function to compute $\chi^2$ accounts for
the correlations in the observed measurements between bands, unlike
the usual assumption that measurements are independent between bands.

The covariance matrix $\mathbb{C}(\theta)$ is a $N \times N$ matrix
and can be conceptually thought of as a combination of three
components such that
\begin{equation}
\mathbb{C}(\theta) = \mathbb{C}_P(\theta) + \mathbb{C}_\mu(\theta) + 
  \mathbb{C}_C(\theta) 
  \label{eq:error_model}
\end{equation}
where $\mathbb{C}_P(\theta)$ is a diagonal covariance matrix of the
photon counting uncertainties\footnote{In the regime of many
  photons.}, $\mathbb{C}_\mu(\theta)$ is the 
covariance matrix due to crowding uncertainties, and
$\mathbb{C}_C(\theta)$ is the covariance matrix giving absolute flux
calibration uncertainties.  Correlations between bands are significant for
the crowding and absolute flux uncertainties
(\S\ref{sec_noise_model}).  It is worth noting that
eq.~\ref{eq:chi2_error} reduces to the standard $\chi^2$ equation
\citep[e.g., eq.~12.11 in][]{Taylor97} for a diagonal covariance
matrix (independent measurements).

Having established our likelihood function, we use Bayes's rule to
write the probability of the model parameters given the observations,
i.e., the posterior probability distribution function (pPDF), as
\begin{equation}
  P(\theta \midline F_D) \propto P(F_D \midline \theta)\,P(\theta) \,
  \label{eq:bayes}
\end{equation}
where the prior $P(\theta)$ reflects any external or additional
independent information placed on the model parameters.  We discuss
reasonable priors and illustrate them in the context of the PHAT
survey in \S\ref{sec_priors}.

Our fitting technique assumes $F_D$ is fully populated with
measurements (e.g., there are no upper limits).  This avoids the
computational complexity of properly accounting for upper limits in the
fitting and produces fitting that reflects the full
measurements.  Modern photometry codes routinely produce flux measurements
in all bands for any detected source \citep[e.g.,][]{Dolphin00,
  Anderson08} removing the burden to account for upper limits in SED  
fitting.

\section{Dust Extinguished Stellar Model}
\label{sec_sed_model}

The physical model used in the BEAST provides predictions of the SED
of a single star extinguished by dust.

\subsection{Single Star Intrinsic SED}
\label{sec:stellarmodel}

The physics of a star's intrinsic SED can be described by a
combination of stellar atmosphere and stellar evolutionary models.
The wavelength dependent luminosity, i.e., its SED, of a single star
with birth mass $M$, age $t$, and metallicity $Z$ is
\begin{eqnarray}
L_{\lambda}(M,t,Z) &=& 4 \pi\, \left[ R(M,t,Z) \right]^2 \nonumber\\
                   &&  \times \, S_{\lambda}(T_{\mathrm{eff}},\log(g),Z ) \, ,
  \label{eq:stellarlum}
\end{eqnarray}
where $R(M,t,Z)$ is the star's radius, given by stellar evolution
models, and $S_{\lambda}(T_{\mathrm{eff}},\log(g),Z)$ is the star's
surface flux given by stellar atmosphere models and parameterized by
the star's effective temperature, $T_\mathrm{eff}$, surface gravity,
$\log(g)$, and metallicity $Z$.  Stellar evolution models provide a
mapping of $(M, t, Z)$ to $(T_{\mathrm{eff}}, \log(g), Z)$,
effectively reducing the number of free parameters needed to fully
describe a star's SED.  We have selected ($M, t, Z)$ as the
fundamental stellar parameters because there is a direct mapping given
by the stellar evolution models
between ($M,t,Z$) and ($T_\mathrm{eff},\log(g),Z)$, whereas the
reverse has degeneracies such that at the same ($T_\mathrm{eff},
\log(g)$) can have multiple possible values of ($M, t$).  Thus, we
write the fundamental stellar parameters as:
\begin{eqnarray}
  \theta_{\mathrm{star}} & = & \{M,t,Z \} \, .
  \label{theta_star}
\end{eqnarray} 

The BEAST uses a merger of two popular stellar atmosphere
grids, the local thermal equilibrium (LTE) CK04 grid
\citep{Castelli04} and non-LTE TLusty OSTAR and BSTAR grids
\citep{Lanz03, Lanz07}.  The merging was done by preferring the TLusty
non-LTE models over the CK04 LTE models in regions of overlap, given
the higher fidelity of the non-LTE modeling for hot stars although the
spectra of stars in the overlap regions are very similar
\citep{Lanz03}.  We confirmed that both grids give very similar
spectra for the same atmospheric parameters in the overlap region as
has been noted previously \citep{Lanz03}.
This provides a seamless merged grid that has excellent ($T_\mathrm{eff},
\log(g)$) coverage, as illustrated by
Fig.~\ref{fig_stellar_mod_overlap}.  The stellar evolutionary tracks
used by the BEAST are the Padova \citep{Marigo08, Girardi10} or PARSEC
\citep{Bressan12, Bressan13} tracks used in the form of isochrones downloaded
from the CMD website with no additional
interpolation.\footnote{http://stev.oapd.inaf.it/cgi-bin/cmd} The
coverage of the stellar atmosphere and evolutionary track models is
visualized two different ways in Fig.~\ref{fig_stellar_mod_overlap}.
Potential future expansions of the BEAST stellar model would be the
addition of stellar atmosphere models to cover a wider range of stars
\citep[e.g.,][]{Westera02, Bergemann12, Rauch13} and using newer stellar
evolutionary tracks as they become available \citep[e.g.,][]{Chen14,
  Tang14}.

\subsection{Interstellar Dust Extinction}
\label{sec_extinction}

\begin{figure}[tbp] 
\epsscale{1.15} 
\plotone{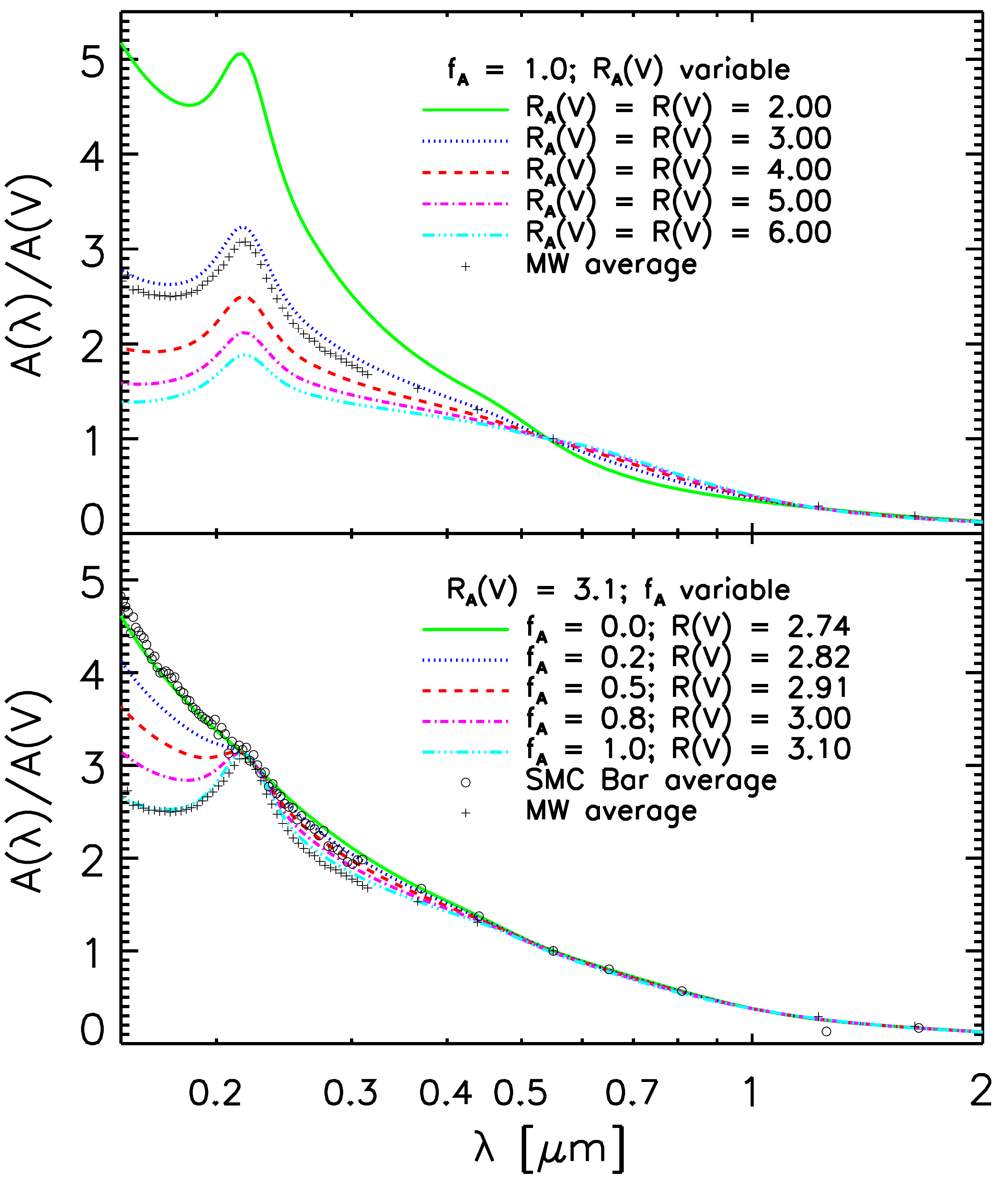}
\caption{The behavior of the new, expanded two parameter ($R(V)$,
  $f_\mathcal{A}$) model for the normalized dust extinction curves is
  illustrated.  The top panel gives the variation in the curves as a
  function of $R_\mathcal{A}(V)$ for the fixed value of $f_\mathcal{A}
  = 1$.  This shows the $\mathcal{A}$ component of the model.
  Overplotted is the average extinction curve for all the Milky Way
  sightlines studied by \citet{Gordon09FUSE}.  The bottom panel gives
  the variation in the curves as a function of $f_\mathcal{A}$ for a
  fixed value of $R_\mathcal{A}(V) = 3.1$.  Overplotted is the same
  Milky Way average curve as in the top panel as well as the SMC Bar
  average extinction curve \citep{Gordon03} that is the $\mathcal{B}$
  component of our model.
\label{fig_new_ext} }
\end{figure}

Interstellar dust extinguishes stellar light as it travels from the
star's surface to the observer.  The wavelength-dependence of the
extinction from the UV to the NIR has been measured along many
sightlines in the Milky Way \citep{Cardelli89, Fitzpatrick99review,
  Valencic04, Gordon09FUSE} and for a handful of sightlines in the
Magellanic Clouds \citep{Gordon98, Misselt99, MaizAppellaniz12} as
well as in M31 \citep[][Clayton et al.\ 2015, submitted]{Bianchi96}.
The observations show a wide range of dust column normalized
extinction curves, $A(\lambda)/A(V)$.  Here we introduce a mixture
model with two components $\mathcal{A}$ and $\mathcal{B}$ to describe
the full range of observed extinction curves in the Local Group.

For most Milky Way sightlines, the variations in dust extinction at a
particular $\lambda$ have been found to be, on average, linearly
dependent on the single parameter $R(V) = A(V)/E(B-V)$
\citep{Cardelli89, Fitzpatrick99review, Gordon09FUSE}.  We adopt this
Milky Way extinction model as the $\mathcal{A}$ component of our
mixture model, expressed as
\begin{equation}
\left[ \frac{A(\lambda)}{A(V)} \right]_{\mathcal{A}} =
   a(\lambda) + \frac{b(\lambda)}{R_{\mathcal{A}}(V)},
   \label{eq:dust_MW}
\end{equation}
where the equations for $a(\lambda)$ and $b(\lambda)$ are determined
from linear fits to $A(\lambda)/A(V)$ versus $R(V)^{-1}$
\citep{Cardelli89}.  We use \citet{Fitzpatrick99review} to compute
$\left[ {A(\lambda)}/{A(V)} \right]_{\mathcal{A}}$ as it is explicitly
formulated to account for passband effects of the optical and NIR
extinction curve measurements.  The behavior of the $\mathcal{A}$
component is shown in the top panel of Fig.~\ref{fig_new_ext}.

Most sight-lines in the Magellanic Clouds do not follow the Milky Way
$R(V)$-dependent relationship \citep{Gordon03}.  For example, the
extinction curves in the Small Magellanic Cloud (SMC) star-forming Bar
lack the usually strong 2175~\AA\ extinction bump and show a
wavelength dependence that is nearly linear versus $\lambda^{-1}$.  We
have found that the Magellanic Cloud and ``deviant'' Milky Way 
sightlines \citep{Mathis92, Valencic04} can be represented by a
mixture model given by
\begin{equation}
k_{\lambda}(R(V),f_\mathcal{A}) =
   f_\mathcal{A} \left[ \frac{A(\lambda)}{A(V)} \right]_{\mathcal{A}} +
   (1 - f_\mathcal{A}) \left[ \frac{A(\lambda)}{A(V)} \right]_{\mathcal{B}},
   \label{eq:dust_mixture}
\end{equation}
where $f_\mathcal{A}$ gives the fraction of the $\mathcal{A}$-type
extinction and $(1 - f_\mathcal{A}$) the fraction of the
$\mathcal{B}$-type extinction (Tchernyshyov \& Gordon, in prep.).  The
wavelength dependence of the $\mathcal{B}$-type extinction is given
using the SMC Bar average UV parameters and optical/NIR data points
from \citet{Gordon03} using the technique of
\citet{Fitzpatrick99review} to smoothly interpolate between band
extinctions in the optical and NIR.  The $J$ and $K$ band values of
${A(\lambda)}/{A(V)}$ were changed from those given by
\citet{Gordon03} to 0.25 and 0.11, respectively, to provide a smooth,
non-negative cubic spline interpolation.  Note that $f_\mathcal{A}$ is
not a direct measure of the 2175~\AA\ extinction feature, but is a
measure of one component of the extinction curve shape variation
across the full wavelength range.  The behavior of this mixture model
is illustrated in Fig.~\ref{fig_new_ext}.

The $R(V)$ of the mixture extinction curve model is
\begin{equation} 
R(V)^{-1} = f_\mathcal{A} R_{\mathcal{A}}(V)^{-1} + 
           (1 - f_\mathcal{A}) R_{\mathcal{B}}(V)^{-1}
\end{equation} 
where we fix $R_\mathcal{B}(V) = 2.74$ \citep{Gordon03}. The range of
observed $R_\mathcal{A}(V)$ is between 2.0 and 6.0 and this results in
the parameter space defined by ($R(V)$, $f_\mathcal{A}$) not being
completely filled, see \S\ref{sec_priors}.

With this dust mixture model, interstellar extinction is included in
our model with the multiplicative term
\begin{eqnarray}
  D_{\lambda, \mathrm{dust}}(\theta_{\mathrm{dust}}) &=& 
        10 ^ {-0.4 \, A(V) \,  k_{\lambda}(R(V),f_\mathcal{A})}
  \label{eq:attenuation}
\end{eqnarray}
that has the three dust parameters
\begin{eqnarray}
  \theta_{\mathrm{dust}} &=& \{A(V), R(V), f_\mathcal{A} \}
  \label{eq:theta_dust}
\end{eqnarray}
where $A(V)$ is the extinction in magnitudes in the Johnson $V$ band
and the $(R(V), f_\mathcal{A})$ parameter combination defines the
shape of the extinction curve.

\subsection{Full SED Model}
\label{sec:fullmodel}

\begin{figure}[tbp] \epsscale{1.25}
\plotone{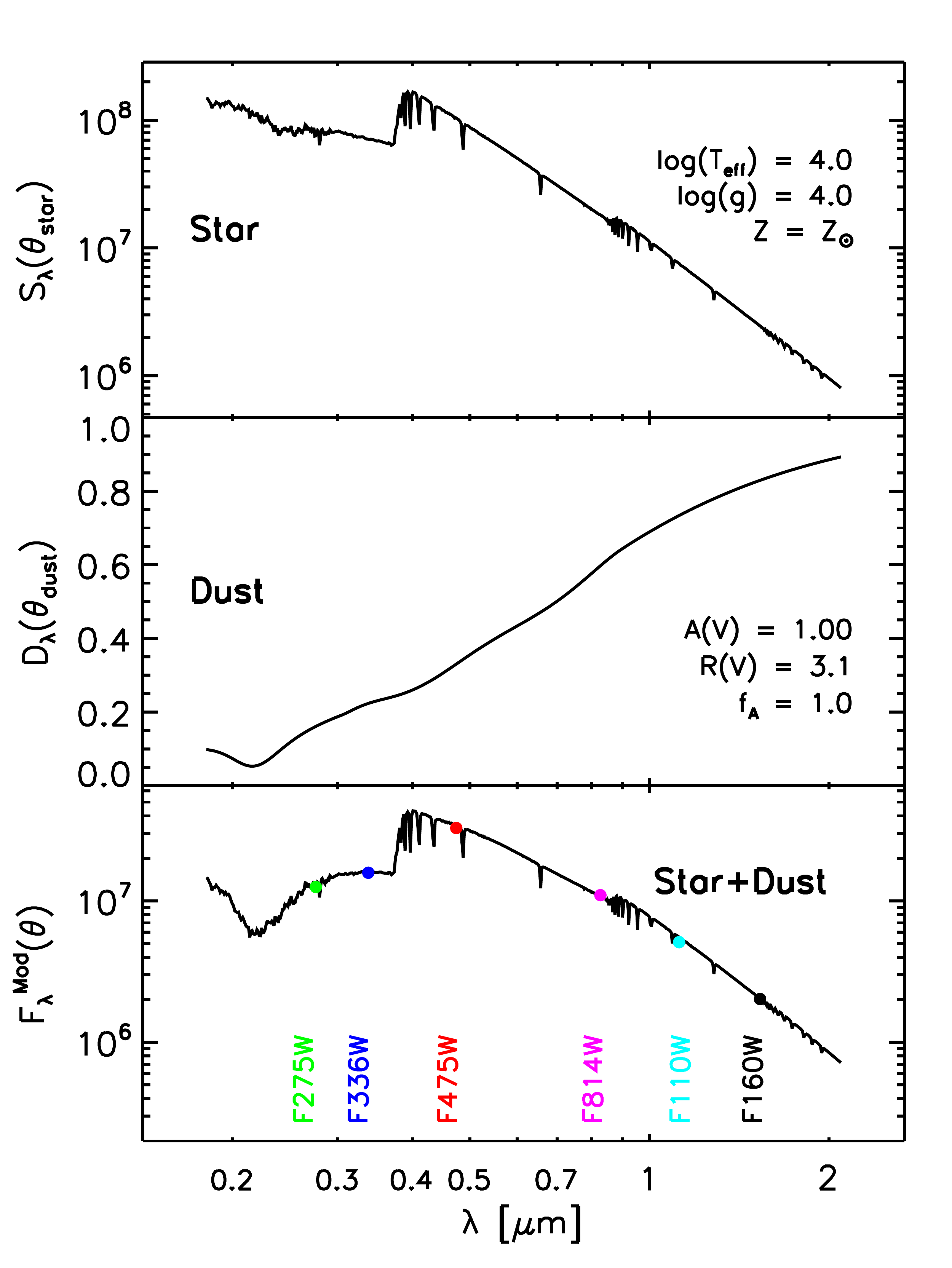}
\caption{The computation of a BEAST model SED for a dust extinguished
  star is shown graphically.  The top panel gives the stellar
  spectrum, the next panel shows the extinction by dust, and the
  bottom panel plots the full extinguished stellar spectrum.  In the
  bottom panel, the integrated SEDs for the PHAT bandpasses are
  plotted at their effective wavelengths ($\lambda_\mathrm{eff}$,
  solid circles).
  \label{fig_exp_spec}}
\end{figure}

The results of the previous two subsections (Eq.~\ref{eq:stellarlum}
and \ref{eq:attenuation}) provide a model of a star's observed
monochromatic flux after passing through a column of dust.  For a
star at a distance $d$, we write the full model for the 
monochromatic flux $F_{\lambda}^{\mathrm{Mod}}$ as
\begin{equation}
\label{eq_model}
F_{\lambda}^{\mathrm{Mod}}(\theta) = \frac {L_{\lambda}(\theta_{\mathrm{star}}) \,
   D_{\lambda, \mathrm{dust}}(\theta_{\mathrm{dust}})}{4 \pi d^2} \,
\label{eq:sed_model}
\end{equation}
where 
\begin{eqnarray}
\theta & = & \set{ \theta_{\mathrm{star}}, \theta_{\mathrm{dust}}, d} \\
  & = & \{M, t, Z, A(V), R(V), f_\mathcal{A}, d \}.
\end{eqnarray}

\begin{figure*}[tbp] 
\epsscale{1.15} 
\plotone{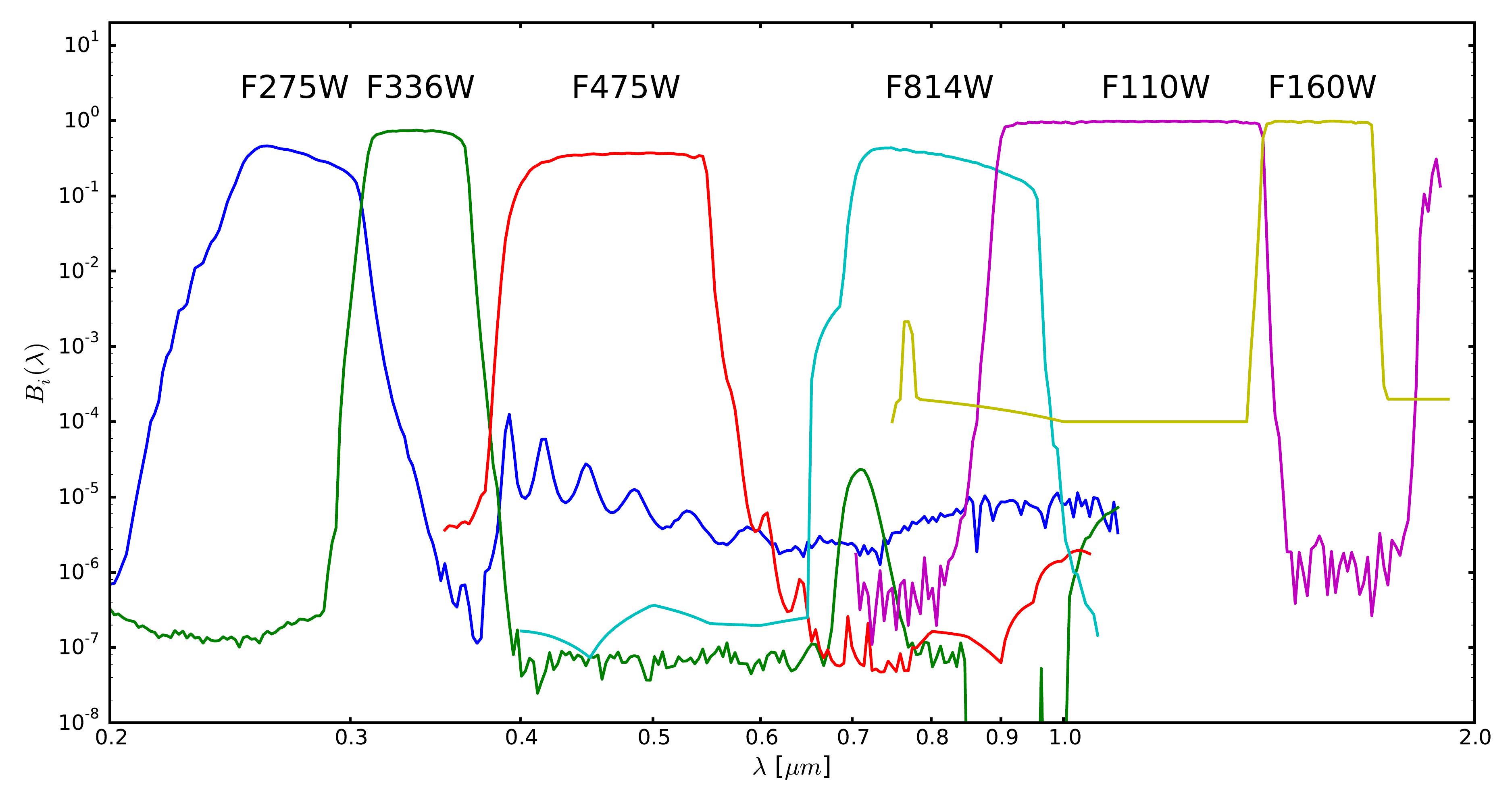}
\caption{The band response functions for the six PHAT filters are
  plotted.  The plot is shown with a log scale to illustrate the
  response seen over the full wavelength range (i.e. showing the red
  and blue ``leaks'').  These response functions include both the
  filter throughputs and the detector efficiencies.
\label{fig_phat_bandpasses}}
\end{figure*}

To compare with photometric observations, we need to compute model
fluxes in the same bands as the observations.  We calculate the model
band flux in bandpass $i$ using
\begin{equation} F_{i}^{\mathrm{Mod}} =
  \frac{\int{\lambda\,B_i(\lambda)\,F_{\lambda}^{\mathrm{Mod}}(\theta) 
\, \mathrm{d}\lambda}}{\int{\lambda\, B_i(\lambda)\,
\mathrm{d}\lambda}} ,
\label{eq:intphot}
\end{equation} 
where $B_i(\lambda)$ is the bandpass response function for the $i$th
band in fractional photon units. This integration is done in photon
units (via $\lambda \, \mathrm{d}\lambda$) to correctly model how the
measurements were made \citep[e.g., photon-based detectors][]{Hogg02,
  Sirianni05}.  We explicitly calculate the flux in each band from the
the response functions and model spectra including dust
extinction, removing the need to deal with color and bolometric
corrections not associated with inaccuracies in the stellar models.

Fig.~\ref{fig_exp_spec} gives a graphical representation of how the
model SEDs are computed showing the intrinsic spectrum, the effects of
dust, the extinguished spectrum, and the band integrated SED.

\section{PHAT Implementation Details}
\label{sec_phat_details}

As a concrete example of the use of the BEAST on a large set of
resolved stellar photometry, we fit the 0.7 million stars detected in
at least 4 bands in Brick 21 of the PHAT survey.  Preparing and
running the BEAST on this data set requires several steps of
implementation, which are described in this section.

\subsection{Grid Implementation}
\label{grid_implementation}

\begin{deluxetable*}{lcccccc}
%\small
\tablecaption{Model Parameters}
%\rotate
\tablehead{   
 \colhead{Parameter} &
 \colhead{Description} &
 \colhead{Min} &
 \colhead{Max} &
 \colhead{Resolution} &
 \colhead{Prior}}    
\startdata  
$\log(t)$ [years] & stellar age & 6.0 & 10.13 &  0.05 & flat SFR  \\   
$\log(M)$ [$M_\odot$] & stellar mass & -0.8 &  2.0 & variable\tablenotemark{a} & Kroupa IMF\tablenotemark{b} \\
$\log(Z)\tablenotemark{b}$ & stellar metallicity & -2.3 & 0.1 & 0.1 & flat \\
$A(V)$ [mag] & dust column &  0.0  & 10.0 & 0.02 & flat \\
$R(V)$ & dust average grain size & 2.0 & 6.0 & 0.5 & peaked at $\sim$3 (Fig.~\ref{fig_priors}) \\
$f_\mathcal{A}$ & dust mixture coefficient & 0.0 & 1.0 &   0.1  & peaked at 1 (Fig.~\ref{fig_priors}) \\
$d$ & distance & \multicolumn{3}{c}{776 kpc} & $\delta$ function 
\enddata
\tablenotetext{a}{Determined by stellar lifetime and fair sampling
of stellar evolutionary phases.}
\tablenotetext{b}{In future work, we will the \citet{Weisz15} updated
  M31 IMF.}
\tablenotetext{c}{$Z$ is the mass fraction of all elements other than
hydrogen and helium.}
\label{tab:parameters}
\end{deluxetable*}

We have implemented the BEAST probabilistic fitting using a grid-based
approach.  This approach ensures that the entire topology of the
posterior function is explored (i.e., the full model parameter space
$\theta$).  For realistic sampling of the model grid this approach is faster than
a Monte Carlo Markov Chain (MCMC) or nested sampling approach.
Finally it avoids the computationally intensive normalizations needed
for MCMC results for further use of the results (e.g., hierarchical
models of stellar clusters, dust geometry, etc.).  For the PHAT
survey, the final range and grid spacing of each parameter is given in
Table~\ref{tab:parameters}.

\subsection{Response Functions}
\label{response_functions}

For PHAT, the observed bands are HST/WFC3 F275W, F336W, F110W, and
F160W and HST/ACS F475W and F814W.  The full bandpass response
functions used in creating the model SEDs are shown in 
Fig.~\ref{fig_phat_bandpasses}.  The bandpass response functions are
for the full instrument plus telescope system including the detector
response.  This figure shows that these
photometric bands can be quite broad.  For such broad photometric
bands, it is important to include the effects 
of dust extinction before 
integrating across the photometric bands, given that the SED spectral
shape changes the effective wavelength of the measurement.

Integrating the dust-free stellar SED and then multiplying by the dust
extinction using a constant $A(\mathrm{band})/A(V)$ would not only be
formally incorrect, but it also leads to large errors in the model
band fluxes.  For example, applying the dust extinction after band
integration for BEAST models bright enough to be formally detected in
the PHAT survey with $R(V) = 3.0$ and $f_\mathcal{A} = 1.0$ 
would result in errors $> 10$\% for 52\%, 1\%, 24\%, 25\%,
24\%, and 0\% 
for the F275W, F336W, F475W, F814W, F110W, and F160W bands,
respectively.  The maximum error for the same models is 56\%, 15\%,
45\%, 34\%, 29\%, and 3\% for the same bands.  These results are
determined by three factors: the largest $A(V)$ detectable at the
survey depth in a band increases with wavelength; the increasing
$A(\mathrm{band})/A(V)$ with decreasing wavelength; and the width of a
band's response function.  Overall, the trend is for the errors to
decrease as the nominal wavelength of the band increases.  The notable
exceptions to these trends are for F336W and F160W which have the
lowest errors as these bands having significantly
narrower band response functions than the other four bands (see
Fig.~\ref{fig_phat_bandpasses}).

\begin{figure}[tbp] 
\epsscale{1.15} 
\plotone{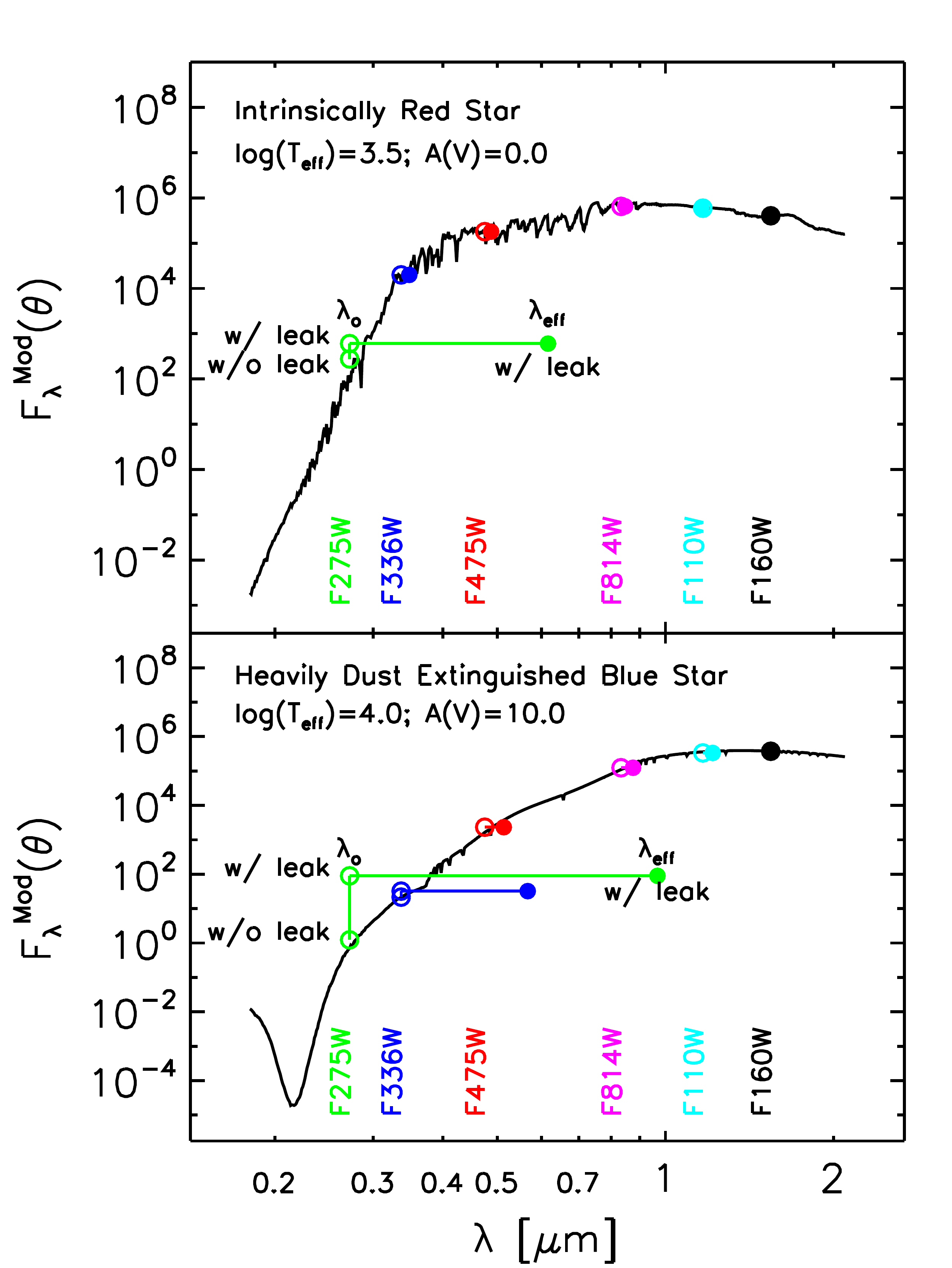}
\caption{The spectra and photometric SEDs for an intrinsically red
  star and a heavily dust extinguished blue star are plotted as
  computed by the BEAST.  The band integrated SEDs are plotted at
  their effective wavelengths ($\lambda_\mathrm{eff}$, solid circles)
  and the nominal wavelengths ($\lambda_o$, open circles).  To
  illustrate the impact of the red leaks, additional open circles
  labeled with ``w/o leak'' for
  the F275W and F336W bands are plotted giving the fluxes computed
  where the red leaks are removed (i.e., zeroing the band response
  functions at $\lambda >$ 0.37 and 0.4~$\micron$, respectively).  The model
  parameters not indicated on the plots are $\log(g)=4.0$, $Z_\sun$,
  $R(V) = 3.1$, and $f_\mathcal{A} = 1$.
  \label{fig_red_leaks}}
\end{figure}

In addition to the broad band response functions, some of the bands
also have red or blue leaks (e.g., F336W with a red leak at
$\sim$0.7~$\micron$, F110W with a strong red leak at
$\sim$1.9~$\micron$, and F160W with a blue leak at
$\sim$0.8~$\micron$).  The impact of such leaks is illustrated in
Fig.~\ref{fig_red_leaks} for an intrinsically red star and a heavily
dust extinguished blue star.  The band integrated fluxes for a number
of the filters are influenced by the steep spectrum inside the main
bandpass and the two bluest filters are strongly influenced by the
contribution from the filter red leak.  The impact of red leaks are
strong for both stars' F275W fluxes and for the F336W flux of the
reddened blue star.  These predictions illustrate that for such red
sources, the bluest filters will collect more photons than nominally
expected due to the red leak signal being much larger than the main
band signal.

One way to quantify the impact of the leaks is to calculate the
effective wavelength of the predicted fluxes.  The effective
wavelength ($\lambda_\mathrm{eff}$) is the flux throughput weighted
average wavelength, while the nominal wavelength ($\lambda_o$) is the
throughput only weighted average wavelength.
The effective wavelengths show that for the
intrinsically red star, the flux measured in what is a UV filter
(F275W) is actually due to optical photons.  For the reddened blue
star, the effect is even larger with the F275W flux dominated by NIR
photons.  The filter fluxes for sources strongly impacted by leaks are
often found below the actual SED at $\lambda_\mathrm{eff}$.  In these
cases, the measured flux can be thought of as an average of flux in the main band
and the leak band that has a much lower throughput than the main band
with an effective wavelength somewhere in between the main and leak bands.

While the examples given in Fig.~\ref{fig_red_leaks} were chosen to be
extreme to illustrate the impact of red leaks, a portion of the
parameter space will be impacted by such leaks at a measurable level.
For the models that have F275W fluxes detectable with the PHAT
observations (i.e., $> 5\times10^{-19}$
ergs~cm$^{-2}$~s$^{-1}$~\AA$^{-1}$), the F275W red leak has a larger
than 5\% effect for 9\% 
of the BEAST models with a maximum impact of $\sim$50\%.  For the
models that have F336W fluxes detectable with the PHAT observations
(i.e., $> 1\times 10^{-19}$~erg~cm$^{-2}$~s$^{-1}$~\AA$^{-1}$), the
F336W red leak has a larger than 5\% effect for 0.6\% of the BEAST
models with a maximum impact of $\sim$14\%.

Thus, correctly performing the filter integrations after fully
generating the model of a dust extinguished star is critical to
achieve high precision and avoid systematic biases in the resulting
model fit parameters.

\subsection{Priors}
\label{sec_priors}

\begin{figure}[tbp] 
\epsscale{1.1} 
\plotone{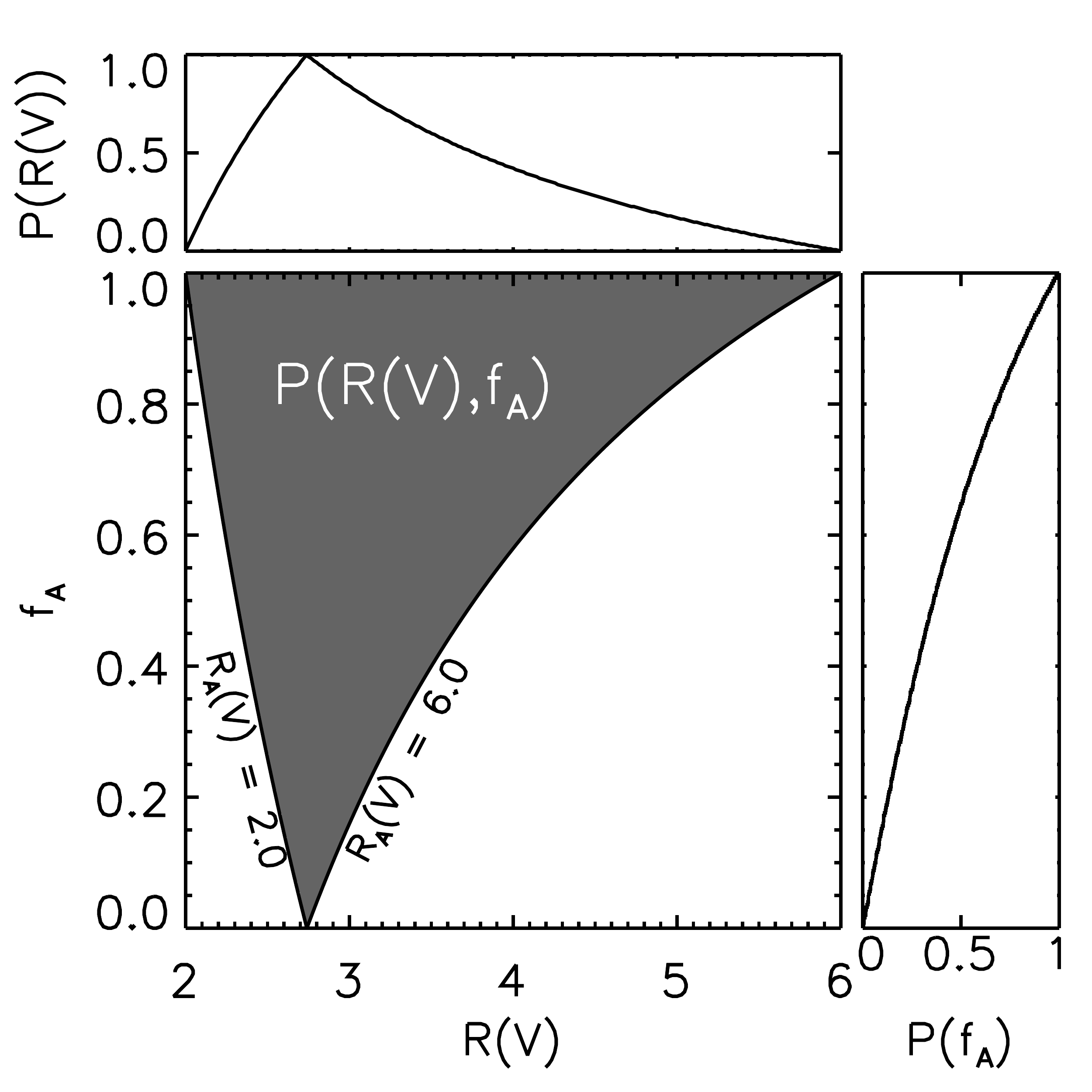}
\caption{The prior on $R(V)$ and $f_\mathcal{A}$ is shown, both in 2D
  and in 1D forms.  The prior is set by imposing a fixed range in
  $R_\mathcal{A}(V)$ from 2 to 6 as this corresponds to the observed
  range for the Milky Way sightlines used to define the $\mathcal{A}$
  component \citep{Cardelli89, Fitzpatrick99review, Gordon09FUSE}.
  \label{fig_priors}}
\end{figure}

All fitting includes priors, whether they are acknowledged or not.
Explicitly using priors on the model parameters provides a clear way to
acknowledge and 
quantify what are often seen as assumptions in the fitting process.
Priors also provide a quantitative way to incorporate 
independent knowledge about the stars and dust in a galaxy from
previous studies in the fitting.  For example, imposing a prior on
the stellar masses 
by using an Initial Mass Function (IMF) can greatly help the
statistical accuracy of the fitting, since the known steep IMF favors
the production of low mass stars \citep[see the review
  of][]{Bastian10}.  Priors have a significant effect when
the data does not provide strong constraints on a parameter.  For
example, priors have a strong impact on 
low signal-to-noise measurements but only a minor impact on high
signal-to-noise measurements.  Given that the BEAST has been developed
to fit survey observations where the majority of detected sources have
a low signal-to-noise, it is important to carefully choose the priors.

The BEAST fitting and marginalization is done using a grid due to the
complexity of the physical model and need for computational speed.  A
uniform grid in all dimensions provides for the easiest and fastest
implementation as the integration needed to marginalize over any
parameter becomes a simple summation as all points have the same nD
volume.  Computational and memory considerations motivate a logarithmic
or non-uniform spacing for some of the parameters.  We account for the
such spacing and impose reasonable and physically motivated
priors using weights.

The priors that result from the final weighting are summarized in
Table~\ref{tab:parameters}.
We adopt a flat prior on $A(V)$, as we have no prior information about
the expected $A(V)$ distribution.  We implement this prior using
equal spacing of the $A(V)$ points in the model grid.  We adopt a
uniform prior in the 2D space of $R(V)$ 
versus $f_\mathcal{A}$ subject to the constraint that $2 <
R_\mathcal{A}(V) < 6$ as shown in Fig.~\ref{fig_priors}.  This
produces a 1D projection of the
prior weighted towards high values of $f_\mathcal{A}$ and $R(V) \sim
3$.  The resulting $R(V)$ and $f_A$ priors are
reasonable as they roughly reflect the range of observed $R(V)$ values
\citep[e.g., Fig.~3 in][]{Gordon09FUSE} and the strong dominance of
$f_\mathcal{A} \sim 1$ curves in the total sample of observed
extinction curves \citep{Gordon03, Valencic04}.

\begin{figure}[tbp] 
\epsscale{1.15} 
\plotone{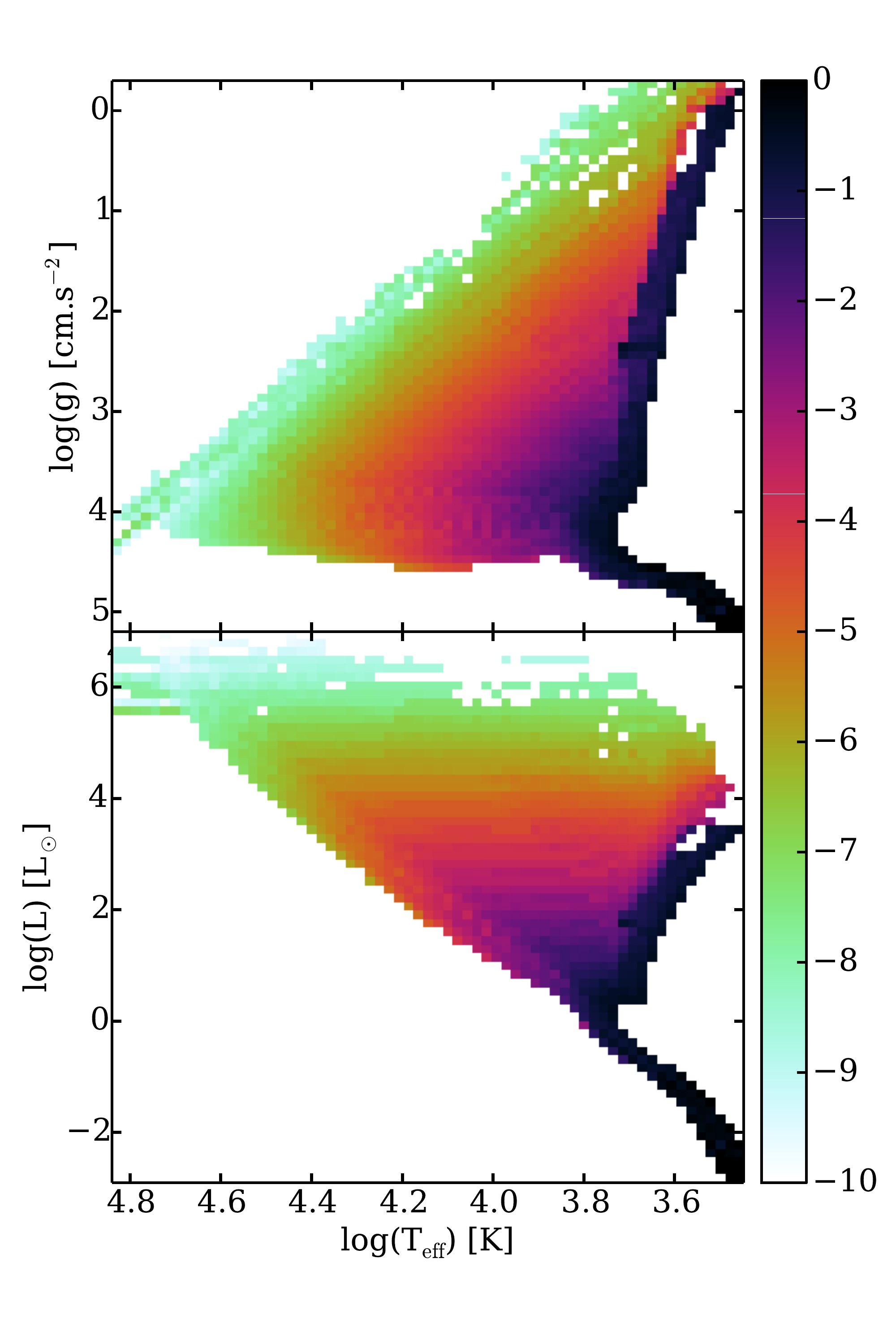}
\caption{The mapping of the stellar $t$ (age) and $M$ (mass) priors
  into the $\log(T_\mathrm{eff})$ versus $\log(L)$ and
  $\log(T_\mathrm{eff})$ versus $\log(g)$ spaces are shown in a log
  scaling with the peak weight set to 1.0.  We use cubehelix color
  scalings for this figure and throughout our paper as such color
  scales are robust for various types of color blindness \citep{Green11}.
  \label{fig_prior_age_mass}}
\end{figure}

The intrinsic grid sampling in stellar mass is driven by the need for
the stellar evolutionary model outputs to efficiently sample the
evolutionary phases.  This sampling is quite different from that of a
reasonable IMF.  We use a multiplicative weight on the grid points to
achieve a Kroupa IMF \citep{Kroupa01} prior at all ages. 
We add an additional multiplicative weight that is a function of age
($t$) to give a uniform prior, as the intrinsic $t$ grid
points are logarithmically spaced for computational speed.  Finally,
we add a third multiplicative term to impose a flat prior on the
stellar metallicity ($Z$) distribution. The mapping of the $M$ and $t$
priors into Hertzsprung-Russell and stellar atmosphere
$\log(T_\mathrm{eff})$ versus $\log(g)$ diagrams is shown in
Fig.~\ref{fig_prior_age_mass}.  The final priors reflect the expected
distribution of real stars, with high densities of low mass stars and
older red giant branch (RGB) stars.

The final prior is on M31's distance for which we assume a value
of 776~kpc \citep[as adopted by][]{Dalcanton12}.  This choice is well justified given
that the distance to M31 is well measured and the high density of M31
stars ensures very small contamination by MW foreground sources.  The
potential variation in the distance to M31 due to the depth of the
galaxy is also small (e.g., a 20~kpc radius is $<$$3\%$ of M31's
distance).  We tested the sensitivity of our results to the assumed
distance.  We find that the recovered parameters (within the 1$\sigma$
confidence intervals) do not change with distance variations on the
order of 10\% for the PHAT bands and survey depth.

\subsection{Noise Model}
\label{sec_noise_model}

The noise model in the BEAST is defined by the bias vector $\mu$ and
the covariance matrix $\mathbb{C}$ which is composed of three terms:
photon, crowding, and absolute flux calibration
(Eq.~\ref{eq:error_model}).

\subsubsection{Absolute Flux Term}

The wavelength dependent absolute calibration of HST is based on the
average of spectroscopic measurements of the predicted to observed
ratios of three white dwarf stars.  The full details are given by
\citet{Bohlin14} and the details relevant to this work are summarized
here.  The   
predicted spectrum for each white dwarf is based on ground-based
spectroscopic measurements of their stellar atmosphere parameters (e.g.,
$T_\mathrm{eff}$, $\log(g)$, etc.).  This predicted spectrum is scaled
by the measured ratio of their fluxes to that of Vega at 5556~\AA\ and
the absolute measurement of Vega's flux at this wavelength compared to
laboratory calibrated blackbodies.
The absolute flux calibration covariance matrix $\mathbb{C}_C$ for HST
has been derived based on uncertainties in the calibration
steps \citep{Bohlin14}.  The HST $\mathbb{C}_C$ is 
composed of two components.  The first captures the uncertainties and
covariance associated with variations in the spectral shape of the
stellar atmosphere models of the three 
white dwarfs (i.e., uncertainties in $\log(T_\mathrm{eff})$,
$\log(g)$, etc.).  The second component is a uniform 0.7\% fully
correlated uncertainty due to the uncertainty in measurements of the
absolute flux scale via Vega's flux at 5556~\AA.   Following the
recommendations by 
\citet{Bohlin14}, we generated the photometric band covariance matrix
for each model by
averaging the spectroscopic resolution matrix using the PHAT filter
response functions and adding in quadrature the 0.7\% uncertainty to
all matrix elements.  The terms of the final absolute flux covariance
matrix are 
\begin{equation}
\mathbb{C}_C^{ij} = F_M^i(\theta) F_M^j(\theta)  \mathbf{A}^{ij}(\theta)
\end{equation}
where $F_M(\theta)$ is the predicted flux in the $N$ photometric bands
(eq.\ref{eq:intphot}) and the $\mathbf{A(\theta)}$ matrix consists of
the fractional uncertainties. 
Fractional values are the natural units of the \citet{Bohlin14} results given the
multiplicative nature of the absolute calibration. We illustrate the
range in the $\mathbf{A(\theta)}$ matrix over the model grid by giving
the minimum and maximum values for each term in the matrix.  Thus, 
\begin{equation}
100 \mathbf{A(\theta)}_\mathrm{min}^{1/2} = \left[
\begin{array}{cccccc}
  0.70 &   0.50 &  0.69 &  0.51 &  0.30 &  0.23 \\
  0.50 &   0.70 &  0.69 &  0.54 &  0.40 &  0.37 \\
  0.69 &   0.69 &  0.70 &  0.69 &  0.67 &  0.67 \\
  0.51 &   0.54 &  0.69 &  0.72 &  0.74 &  0.74 \\
  0.30 &   0.40 &  0.67 &  0.74 &  0.83 &  0.85 \\
  0.23 &   0.37 &  0.67 &  0.74 &  0.85 &  0.88 \\
\end{array} \right]
\end{equation}
and
\begin{equation}
100 \mathbf{A(\theta)}_\mathrm{max}^{1/2} = \left[
\begin{array}{cccccc}
  1.20 &   1.11 &  0.75 &  0.79 &  0.83 &  0.85 \\
  1.11 &   1.05 &  0.74 &  0.79 &  0.83 &  0.85 \\
  0.75 &   0.74 &  0.71 &  0.70 &  0.70 &  0.70 \\
  0.79 &   0,79 &  0.70 &  0.76 &  0.79 &  0.80 \\
  0.83 &   0.83 &  0.70 &  0.79 &  0.84 &  0.86 \\
  0.85 &   0.85 &  0.70 &  0.80 &  0.86 &  0.89 \\ 
\end{array} \right] ,
\end{equation}
where we have multipled $\mathbf{A}$ by 100 for display purposes.
The diagonal terms show that the overall level of absolute flux
uncertainty is around 1\%.  The non-diagonal terms show that this
uncertainty is highly correlated between bands.  The minimum and maximum
values show variations around a factor of two at shorter wavelengths
and much smaller variations in the longer wavelengths.  The difference
in variations between shorter and longer wavelengths is directly
related to larger variations in spectral shape between these
wavelength regions.

\subsubsection{Combined Photon and Crowding Term}
\label{sec_asts_term}

The photon and the crowding terms are
combined and measured simultaneously through the use of artificial
star tests (ASTs).  These tests accurately capture the non-linear
interaction between the photon and crowding noise.  The crowding not
only impacts the noise, but also can systematically bias the flux of a
source.  Such systematics become dominant in the case of high stellar
crowding, when a star's flux measurement is contaminated by a
non-negligible amount of flux from neighboring sources.

ASTs are performed by inserting artificial stars with known SEDs into
the observed images and re-running the full photometric pipeline
on the altered images. 
In this way, the fluxes of the artificial stars are extracted using
the same technique that is used to produce the observed point source
catalog and can be compared to the true input fluxes.

We use the model grid SEDs as inputs for the ASTs.  The
photon noise is simulated using the parameters of the observations
(sensitivity, exposure time, etc.) and each star is inserted into the
observed images using the known point spread function.  Finally, each
star is either measured or undetected as if it were a real source.
This process is repeated many times to build up
secure statistical information on how photometric measurements are
affected by crowding, photon noise, brightness, etc.

The bias
$\mu_i(\theta)$ in band $i$ given $N$ ASTs recovered fluxes $F_R^k$
with an input model $\theta$ that has intrinsic flux $F_M (\theta)$ is
\begin{equation}
\mu_i(\theta) = \frac{1}{N} \sum_k \left( F_{R,i}^k - F_{M}(\theta) \right),
\end{equation}
and the covariance between bands $i$ and $j$ is
\begin{equation}
  \mathbb{C}_{ij}^2 (\theta) = \frac{
    \sum_k^N \left( \mu_i^k - \left< \mu_i (\theta) \right> \right) 
    \left( \mu_j^k - \left< \mu_j (\theta) \right> \right) }
  {N-1}.
\end{equation}
The bias captures the mean offset between the true and measured fluxes
The covariance captures the
magnitude and shape of the scatter about this mean offset.  If the different
bands are independent of each other, as is often assumed, the entries
along the diagonal of the covariance matrix are the squared standard
deviations in each band and all off-diagonal entries are equal to
zero.

\begin{figure*}[tbp] \epsscale{1.2}
\plotone{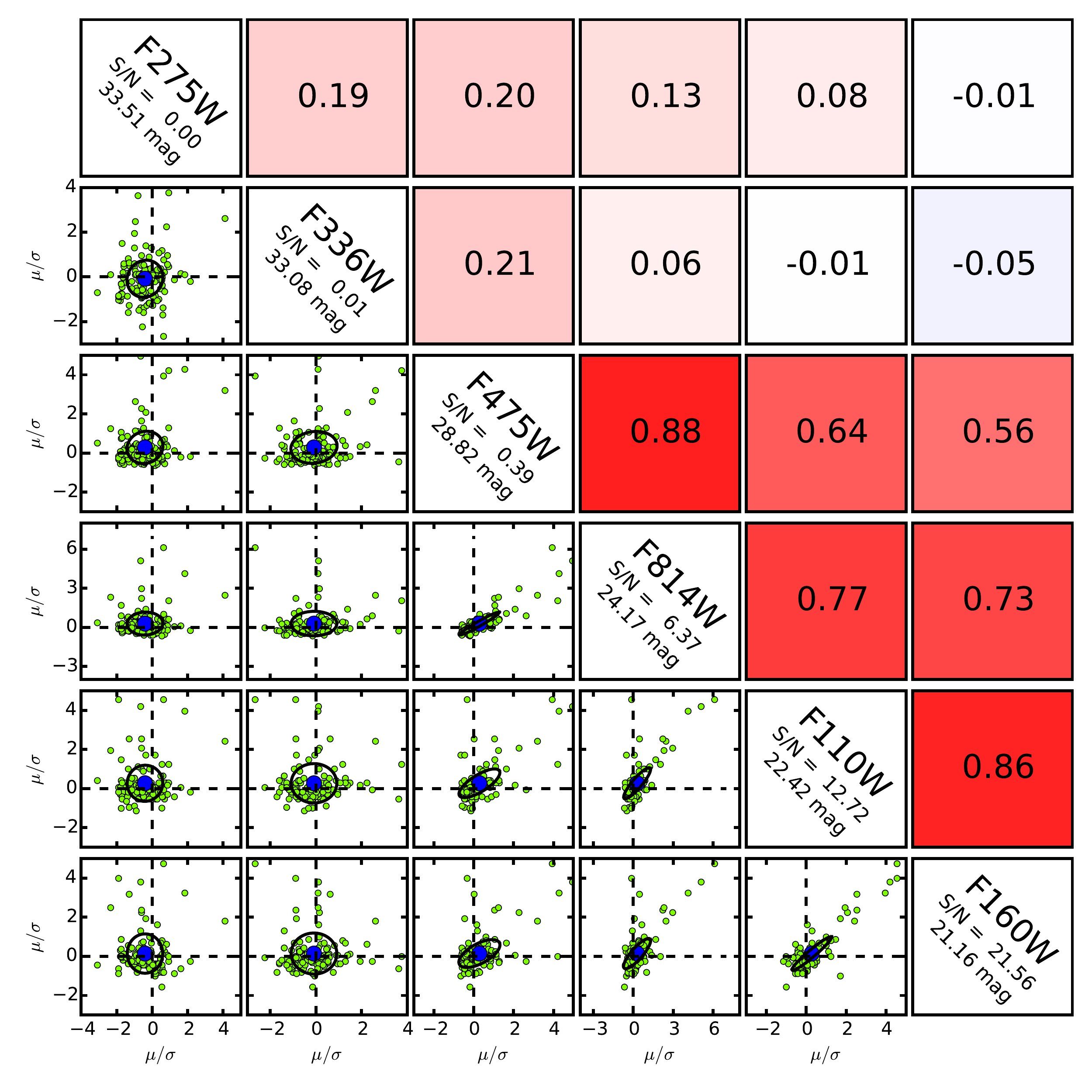}
\caption{The covariance matrix calculation and result is graphically
  illustrated for a single model SED using 125 ASTs with independent
  realizations of the photon noise and spatial location in a small
  region of the PHAT
  observations. The lower triangle of plots shows the individual
  biases $\mu_k$ for each of the $k$ ASTs as small green circles and
  the average bias as a larger blue circle.  The axes of these plots
  are the $\mu_k$ divided by the appropriate band $\sigma_k$. The dashed
  lines give the location of zero bias $\mu_k$ in each band.  The
  circle is an illustration of the covariance calculated from the
  individual $\mu_k$ values.  The upper triangle gives the correlation
  coefficient ($c_ij = \mathbb{C}_{ij}^2/\sigma_i \sigma_j$) for each
  pair of bands.
  \label{fig_cov_ast}}
\end{figure*}

In Fig.~\ref{fig_cov_ast} we show the AST data and
the resulting covariance matrix and bias for a single model SED.
This model SED represents a typical PHAT source
that is well detected in the redder bands and basically undetected in
the three shorter wavelength bands.  The highest covariances seen are
for F475W, F814W, F110W, and F160W bands.  The obvious source of the
covariance is neighboring sources (i.e., crowding noise).  The impact
of crowding noise can be seen even in the F475W band that is detected,
on average, only at $<$1$\sigma$, yet the F475W and F814W bands
are still strongly correlated.  In contrast, the lack of a
strong correlation between the two shortest wavelength bands (F275W
and F336W) and the redder bands is an indication that the measurements in
these bluer bands are dominated by measurement (photon) noise.  

\begin{figure*}[tbp] \epsscale{1.2}
\plotone{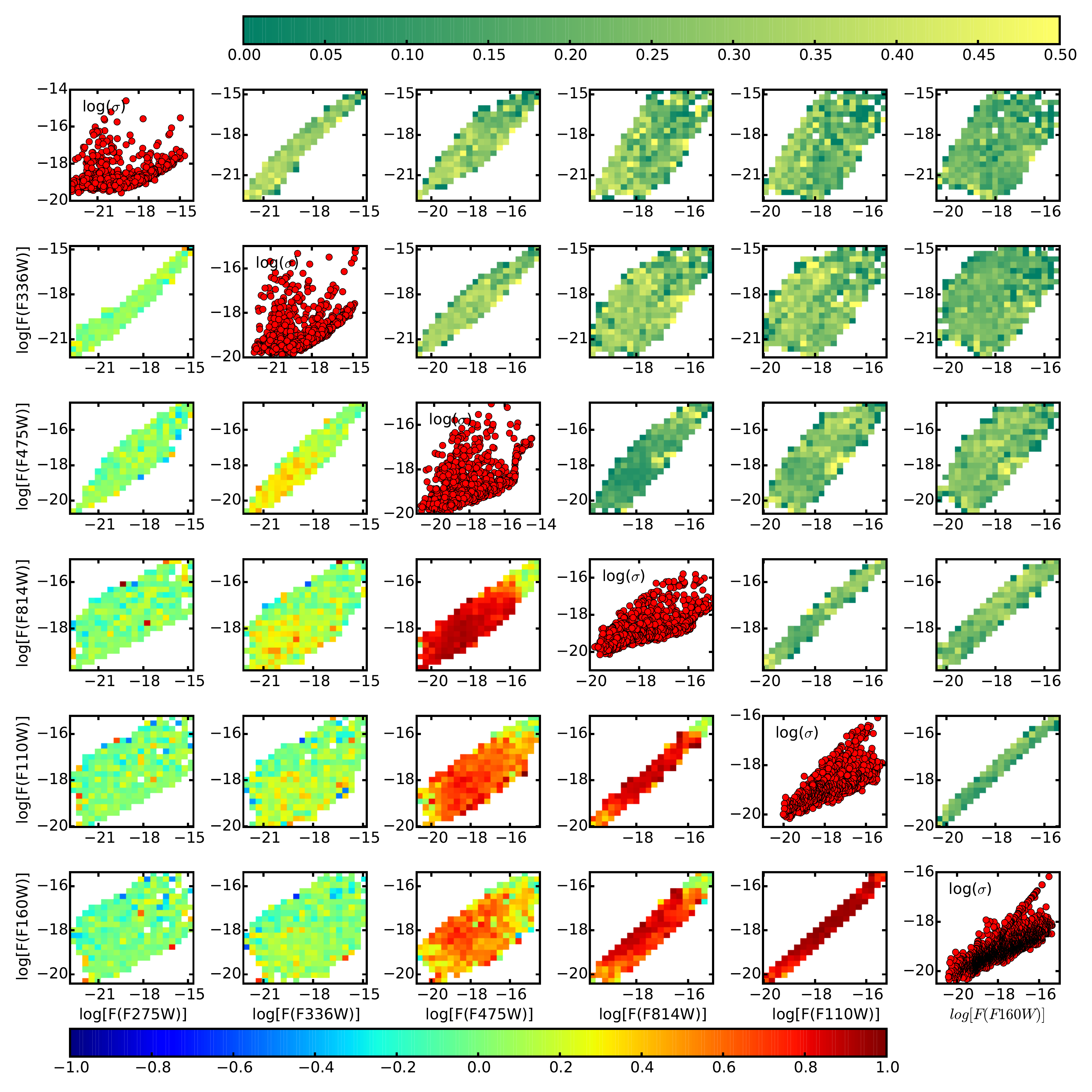}
\caption{A visualization of the covariance matrices is shown for
  Field~2 of PHAT Brick~9.  The plots on the diagonal give
  the $\sigma_k$ values for each band (square root of the diagonals of
  the covariance matrx).  The plots in the lower triangle (color scale
  from lower color bar) give the
  average correlation versus flux in each of the two considered
  bands.  The correlation values are shown instead of covariance
  values to provide a more understandable interpretation of the
  values.
  The plots in the upper triangle (color scale from upper color bar)
  give the standard deviation 
  in the correlation for each of the two considered bands in each
  combined flux bin.  These plots provide a measure of the additional
  variation that is due to the fluxes in other bands than the two
  considered bands.
  \label{fig_all_cov_ast}}
\end{figure*}

Fig.~\ref{fig_cov_ast} also illustrates that our use of a
multi-variate Gaussian to characterize the offset and scatter in the
recovered fluxes is an approximation to the true distributions.  This
can be seen by the asymmetry in the $\mu_k$ values 
in many of the bands with more positive deviations than
negative deviations.  In the future, we will investigate the use of
multi-variate skew Gaussians or numerically sampling the AST
distribution as a refinement of our noise model.  
Practical concerns on the computation cost of including enough ASTs per
model SED to define the deviations from a multi-variate Gaussian may
dominate the discussion.  In practice, we expect that the gains in
including the asymmetries in the $\mu_k$ values via a more
sophisticated noise model wil be smaller than the impact of including
covariance to the first order.

\begin{figure*}[tbp] 
\epsscale{1.15} 
\plotone{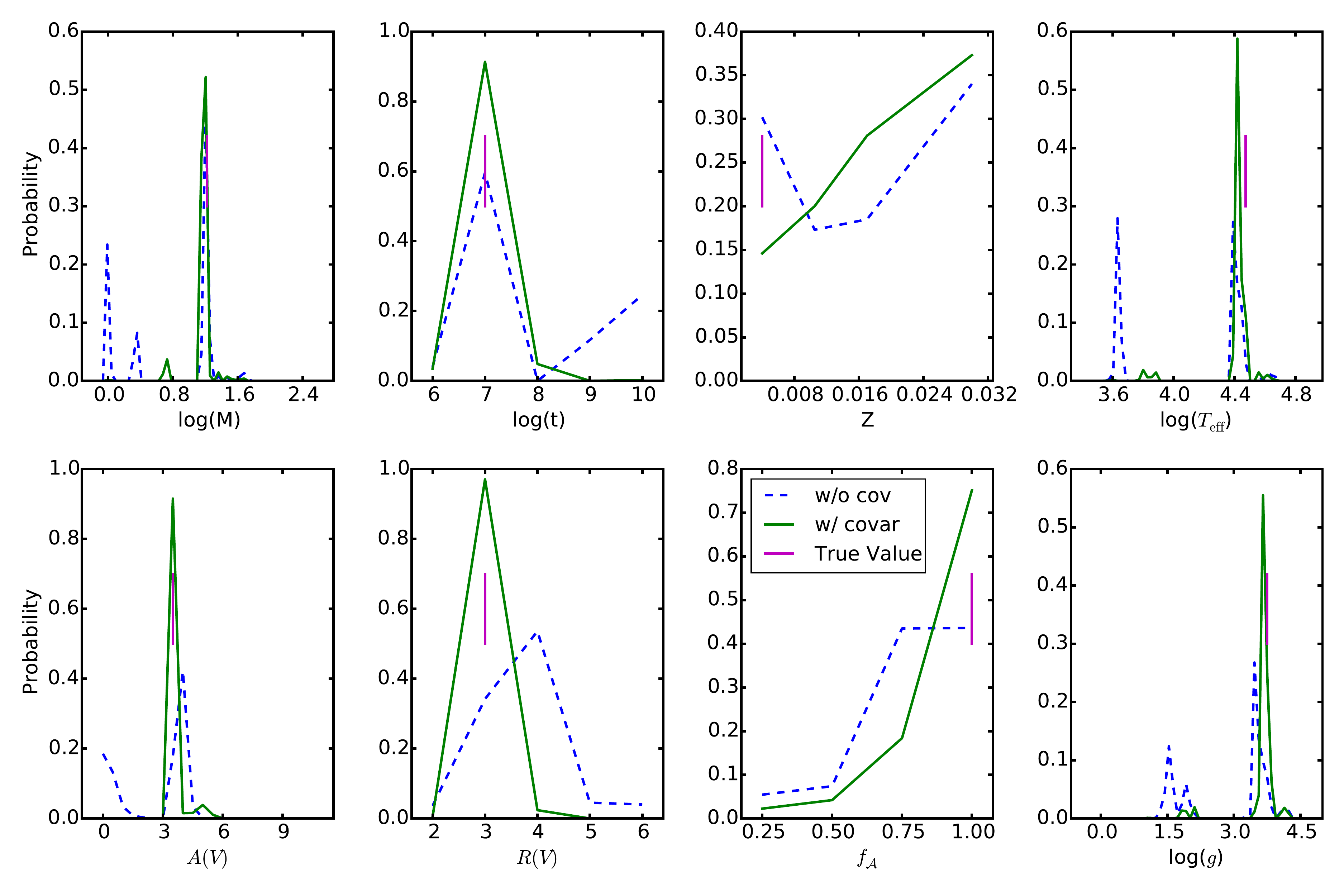}
\caption{The 1D pPDFs are shown from fitting a simulated source as if
  it were observed in Field~15 of Brick~21 of the PHAT survey.  The
  star was simulated using the full noise model including covariance.
  Two sets of 1D pPDFs are shown, one a simple noise model without any
  covariance information and one using the full covariance noise
  model.  The true values of the simulated source are shown.  
  \label{fig_cov_important}}
\end{figure*}

The AST derived covariance matrices and biases are dependent on the
model SED and on the location of a star in the PHAT survey area.
Ideally many ASTs would be run for every BEAST model SED for every 
pixel (or subpixel position) over the entire survey region.  As this
is not computationally feasible, we are forced to average the AST
results over spatial regions and interpolated between models.  The effects of
crowding are most strongly dependent on the source density and, thus,
we average the AST results over regions of similar source density.

We illustrate the dependence of the AST derived covariance matrices in
Fig.~\ref{fig_all_cov_ast} for one field in one PHAT Brick.  This
figure gives projections of the ensemble of the covariance matrices.
The
uncertainty per band is strongly dependent on the source flux as
expected.  The correlation between bands is also strongly dependent on
flux; with fainter fluxes showing stronger correlations, especially in
the longer wavelength bands where crowding is more significant.  The step
function drop in correlation at the highest fluxes is traced to the
fluxes for the brightest stars coming from the the short ``guard''
exposures, rather than the deeper main survey exposures where the
stars are saturated.

\subsubsection{Importance of Including Covariance}
\label{sec_import_cov}

The importance and impact of including the covariance in the SED
fitting is illustrated in Fig.~\ref{fig_cov_important}.  This figure
shows the 1D pPDFs of the SED fitting likelihood function for a
simulated star in Field~15 of PHAT Brick~21.  The simulated source is
a hot, young star that has experienced appreciable dust extinction.
The observational noise was simulated using the full noise model with
a covariance matrix interpolated from those measured using a small set
of full 6-band ASTs spanning the observed flux range run for this
field and brick.  The SED fitting was done twice, the first time
without using the covariance information (i.e., diagonals only) and
the second time using the full covariance matrix.  It is clear that
including the full noise model with covariance produces a more
accurate and precise recovery of true model parameters.  Overall, the
1D pPDFs with covariance better recover the input model parameters,
given that their peaks are better matched to the input values 
and their widths are narrower.  The differences can be dramatic like
those for the $\log(M)$, $\log(T_\mathrm{eff})$, and $\log(g)$ model
parameters where the 1D pPDFs change from being double peaked without
covariance to being dominated by a single peak with covariance.  In
essence, the covariance restricts the allowed parameter space
producing narrow pPDFs.

\subsection{Speed Optimizations}
\label{sec_optimize}

\begin{figure*}[tbp] 
\epsscale{1.1} 
\plotone{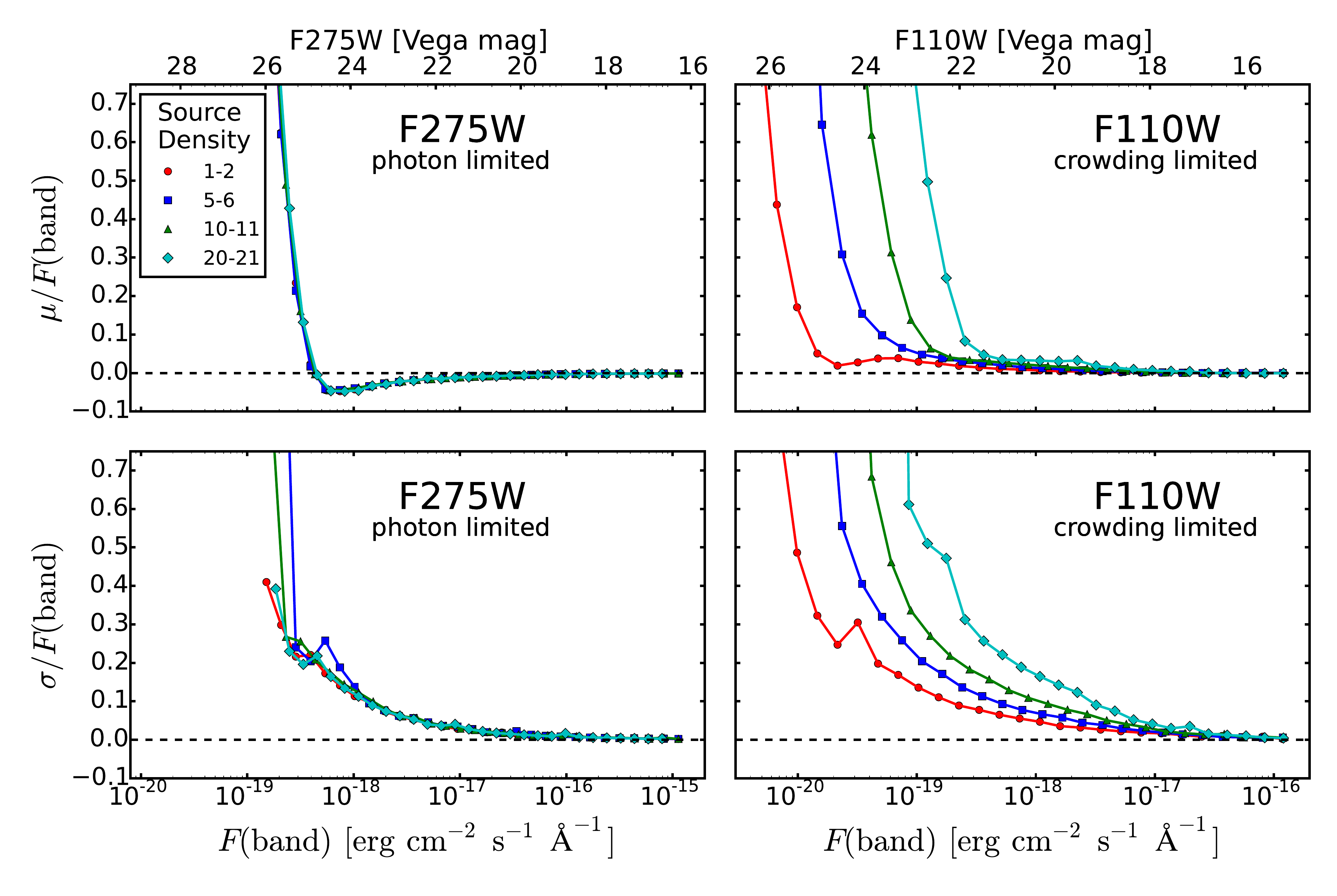}
\caption{The bias ($\mu$) and uncertainty ($\sigma$) terms for the
  F275W and F110W filters are plotted.  These results were calculated
  from ASTs performed over the PHAT survey region.  The source
  densities are in units of stars per $\sq\arcsec$ and 
  were measured in $5\arcsec \times 5\arcsec$ pixels using
  the PHAT sources detected in four bands with F475W magnitudes
  between 24.5 and 27.  This range was chosen as the
  PHAT survey is complete for these brightnesses.  The F275W
  results show the signature of photon limited observations where the
  behavior is independent of source density.  The F110W results
  are strongly dependent on source density which is the signature of
  crowding limited observations.
  \label{fig_noise_vs_sd}}
\end{figure*}

It is fairly quick using modern computers to calculate the full
$N$-dimensional pPDF for a single observed SED using a grid consisting
of millions of models.  The challenge is doing this for the $>$100
million sources in the PHAT survey \citep{Williams14}.  We are
pursuing a number of options for speeding up the pPDF calculations,
mainly focused on reducing the effective size of the model grid.  One
straightforward optimization that we have implemented is to only
compute the pPDF for models that are in the range of fluxes expected
for the survey.  Thus, the grid is trimmed of models that predict
fluxes that would saturate in the survey observations or are well
below the survey sensitivity as measured from the AST results (i.e.,
with zero completeness).  The next potential optimization will be to
compress/hash the grid (similar to a tree-code approach) so that the
BEAST only computes the pPDF for 
models with equivalent SEDs within some tolerance (Fouesneau et al.,
in prep.).  We expect this will allow for faster computation rates
and/or larger model grids.

\section{Example PHAT Results}
\label{sec_phat_results}

To illustrate the BEAST capabilities on real data, we have fit the 0.7
million sources that were detected in at least 4 PHAT bands with a
F475W Vega magnitude brighter than 27.6~mag in the PHAT Brick 21
region \citep{Williams14}.  Brick 21 samples a range of star formation
regions and dust contents.  For this initial work, we use a coarser
grid than given in Table~\ref{tab:parameters}, specifically with
resolutions of 0.15 in $\log(t)$, 0.15 in $A(V)$, 0.5 in $R(V)$ and
0.5 in $f_\mathcal{A}$ and $Z$ = 0.03, 0.019, 0.008, \& 0.004.  This
coarser grid allowed the fitting to be done for the 0.7 million Brick
21 sources in a reasonable amount of time.  With this model grid, we
find that it takes the BEAST $\sim$13~s to fit and save the results
for a single star using the Texas Advanced Computing Center (TACC)
Stampede supercomputer with time provided through XSEDE
\citep{Towns14}.

The noise model was computed using the artificial star tests (ASTs)
discussed by \citet{dalcanton2012b} and \citep{Williams14}.  Using
this noise model, the
uncertainty ($\sigma$) and bias ($\mu$) terms as a function of source
density are shown in
Fig.~\ref{fig_noise_vs_sd} for two filters to illustrate the different
behaviors between photon and crowding limited observations.  This
figure clearly illustrates the dependence of the bias and uncertainty
on flux and source density and how it varies between bands in the PHAT
survey.

The ASTs used in construction of this noise model
were not done simultaneously in all 6 bands, but were instead
done in the pairs of bands associated with each camera due to
computational resource limitations.  Thus, we have conservatively
assumed the AST results are independent between all 6 PHAT bands.
While we have shown that a noise 
model built from the full 6 band ASTs allows the BEAST to provide the
best constraints on the model parameters (\S\ref{sec_asts_term} and
\ref{sec_import_cov}), for the remainder of this study we use the
noise models built from the single camera ASTs to show an example of
the BEAST applied to a large catalog.  While the single camera ASTs do not
contain sufficient information for full covariance measurements, they
do provide good measurements of the dependence of the flux bias and
uncertainty in all 6 bands for the full range of fluxes and crowding
found in the PHAT survey.

\begin{figure*}[tbp] 
\epsscale{1.3} 
\plotone{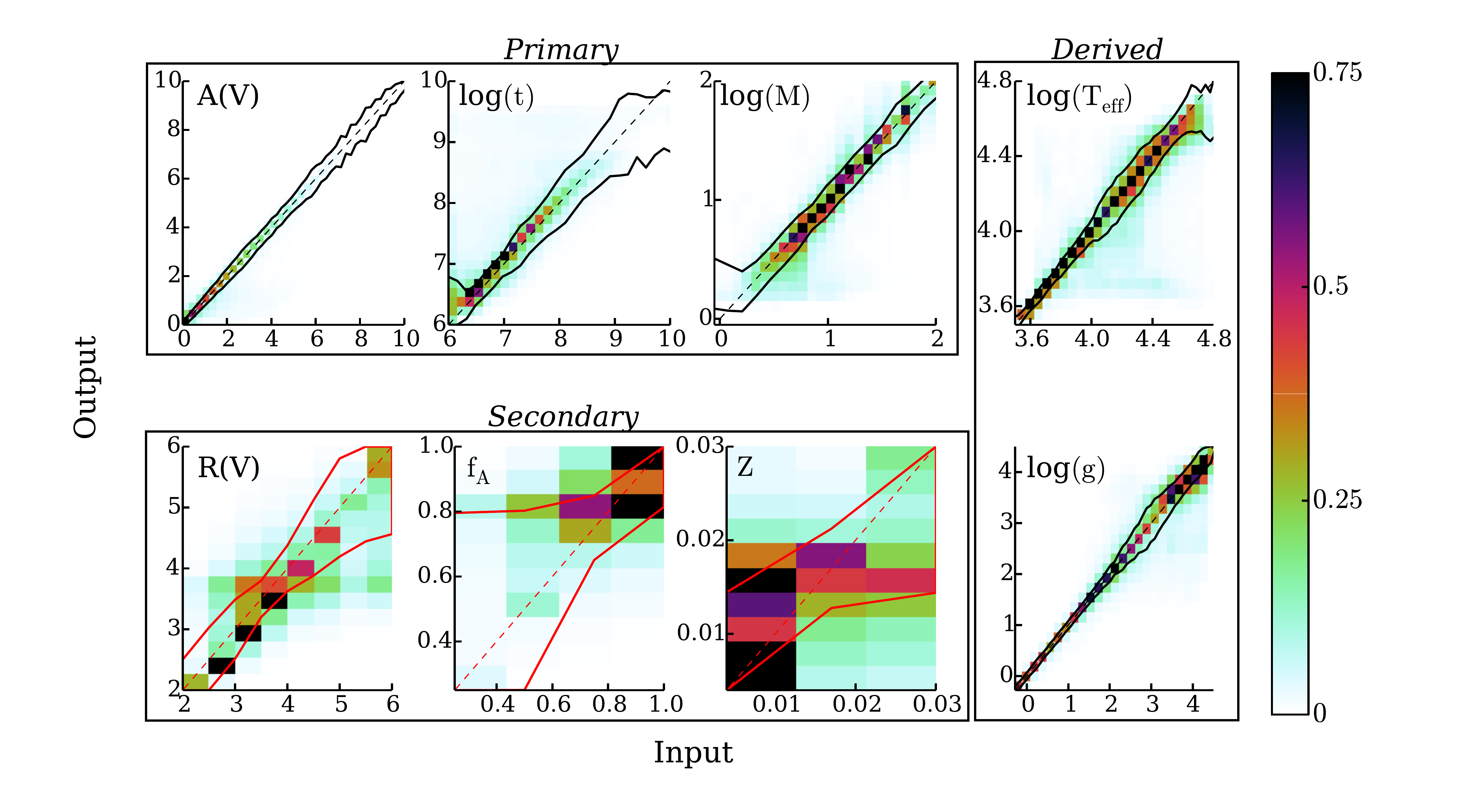}
\caption{The results of sensitivity tests for sources sampled from the
  model grid with the PHAT noise model are shown as normalized density
  plots with a log scaling.  The contours give the 1$\sigma$
  uncertainty (67\%) regions illustrating that the primary and derived
  parameters are well recovered overall.  The precision of the
  recovery of the secondary parameters is lower than for the primary
  parameters as they
  are more strongly influenced by small changes in the SED shapes.
  The density normalization is for display purposes.  The
    distribution of sources is a reflection of the recovery and the
    density of simulated sources which itself is a function of the
    model priors and PHAT observation sensitivity.
  \label{fig_sense}}
\end{figure*}

The assumption of band independent ASTs will lead us to overestimate the
fit parameter uncertainties (wider pPDFs) as assuming independent band
measurements (no covariance) provides the loosest restriction on the
allowed parameter space by the observed SED (see Sec.~\ref{sec_import_cov}).
In the future, new 6 band ASTs will be computed that will provide the
BEAST with a noise model that includes the important covariance
between bands due to crowding.  In all
the following figures, we use the fit parameter expectation values
(i.e., pPDF weighted averages) as these reflect the full range of
allowed models for each source.  This is in contrast to the more usual
use of best fit parameter values that only sample the peak of the
probability distribution function.

\subsection{Parameter Sensitivities}
\label{sec:sensitivities}

We tested how well we can recover the model parameters using the PHAT
survey observations with sensitivity tests.  The results of
sensitivity tests are closely coupled to the assumed noise model.  We
have chosen to use the noise model from PHAT for Brick~21 to provide
realistic sensitivity results using the full PHAT observational
strategy, data reduction, and source extraction.  For these tests,
almost one million models were picked from the set of all
PHAT BEAST grid models with F475W Vega magnitudes brighter than 27.6
mag.  The distribution of sources were picked by randomly sampling the full
prior distribution on the BEAST grid (\S\ref{sec_priors}) modified by
F475W mag limit.  This
provides a reasonable star formation history, distribution of dust
parameters, and PHAT survey sensitivities.  Noise was added to these
models using the source density 
dependent PHAT noise model (\S\ref{sec_phat_results}).  Thus, these sensitivity
tests are directly applicable to the expected BEAST results for real
sources in Brick~21. The resulting recovery of models parameters is
graphically shown in Fig.~\ref{fig_sense}.

These tests clearly separate the parameters into primary ones that
drive the overall SED shape and secondary ones that provide smaller
modifications to the overall shape.  The primary parameters are
$A(V)$, $\log(t)$, and $\log(M)$ and systematically recovered well,
except for sources that have $\log(t) \geq 9$ and $\log(M) \leq 0.3$.
This systematic error in recovery can be traced to the strong
degeneracies between RGB and Red Clump stars for all masses between 1
and 2~$M_\odot$.  The random uncertainty in the recovered parameters
is approximately 0.5 mag, 0.5, and 0.2 for $A(V)$, $\log(t)$, and
$\log(M)$, respectively.  These are averages over the entire parameter
range and specific ranges can be better recovered (e.g., low $A(V)$
values are better recovered than the average).

The secondary parameters are $R(V)$, $f_\mathcal{A}$, and $Z$.  They
are recovered less well than the primary parameters as can be clearly
seen from Fig.~\ref{fig_sense}.  These secondary parameters modify
details of the SED and are more strongly affected by noise than the
primary parameters.  Of the secondary parameters, $R(V)$ is recovered
the best with mild systematics at high $R(V)$ values and an average
random uncertainty of approximately 1.0.  The recovery of
$f_\mathcal{A}$ and $Z$ is fairly poor with systematic offsets and
large random uncertainties of approximately 0.25 and 0.013 for
$f_\mathcal{A}$ and $Z$, respectively.

\begin{figure*}[tbp] 
\epsscale{1.15}
\plotone{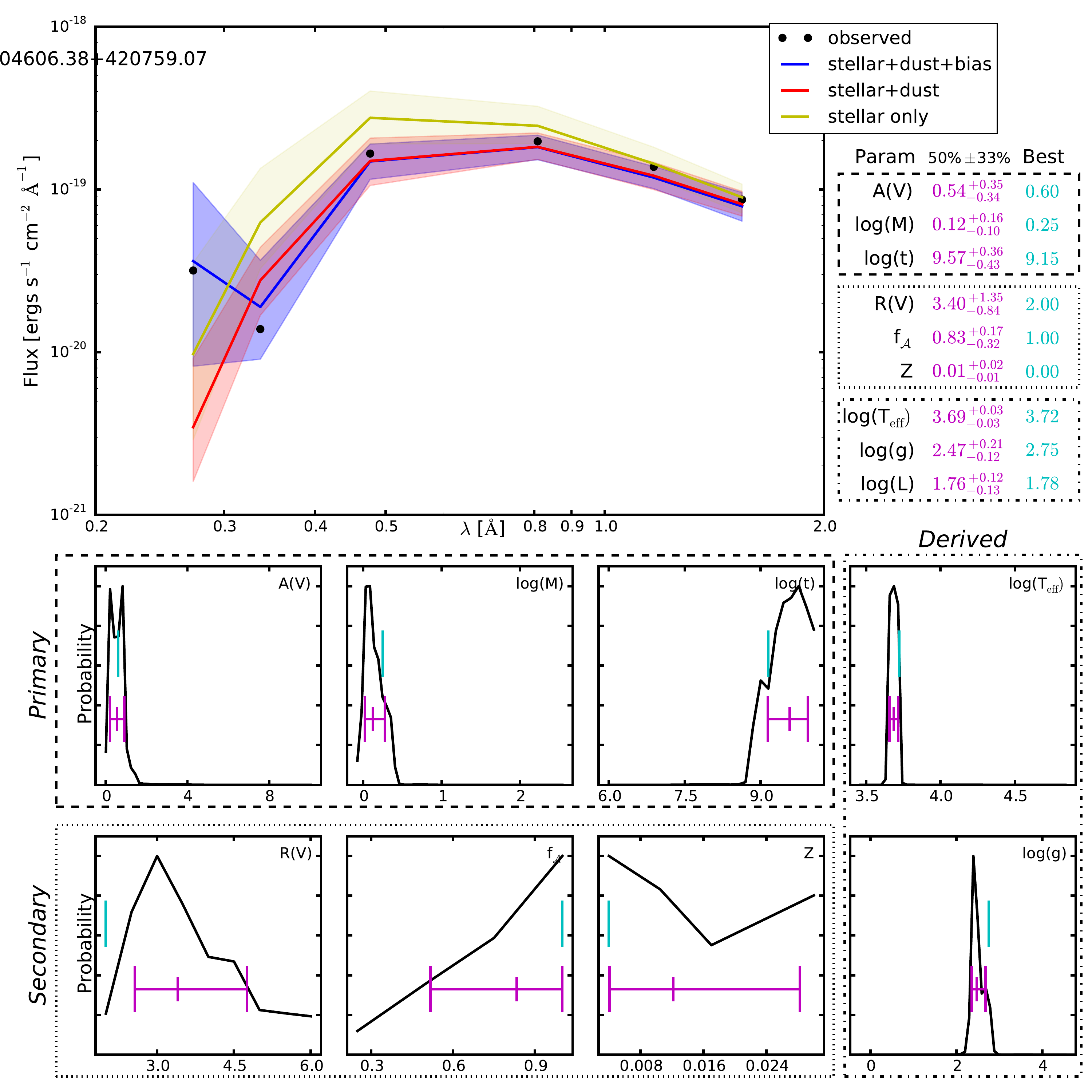} 
\caption{The fitting results for an individual star in Brick~21 are
  shown.  While this figure illustrates the fit to an individual
    source, it is for a red giant star that is one of the common sources
    we find in the PHAT region.
    The 50\% model is shown along with shaded
  colored regions indicating the range of models that fit within
  1$\sigma$.  The models shown are the full model including
  the observational bias
  (stellar+dust+bias), the physical model alone (stellar+dust), and
  the stellar model alone (stellar).  The impact of the
  bias in the photometric measurements is clearly seen in the UV and
  illustrates the importance of a full physical and observational
  model for this type of fitting. 
  The best fit parameter values (cyan) and  $50\% \pm 33\%$ values
  (purple) are given numerically as well as graphically in the 1D pPDF
  plots.  The 1D pPDF plots give the normalized probabilities with the
  y-axis scale ranging from 0 to 1.1.  The best fit values are
  determined from the full 6D pPDFs and, as such, often do not appear
  at the peak of the 1D pPDFs.  
  \label{fig_indiv_sed_fit}}
\end{figure*}

The recovery of the derived parameters $\log(T_\mathrm{eff})$ and
$\log(g)$ is quite good and is similar to the primary parameters.
These parameters are labeled as derived as 
they can be directly derived from the primary parameters $\log(t)$ and
$\log(M)$ at a given $Z$.  We include the derived parameters here as
they are more often used in SED modeling than $\log(t)$ and $\log(M)$
and are potentially easier to use to answer specific science questions
(e.g., investigating the UV radiation field, Kapala et al., in prep.).
They are recovered well systematically over their entire range.  The
random uncertainties are approximately 0.1 and 0.25 for
$\log(T_\mathrm{eff})$ and $\log(g)$, respectively.  These results
show that the BEAST results are not strongly affected by the usually
seen degeneracy between $\log(T_\mathrm{eff})$ and $A(V)$.  This is
due to the wide wavelength coverage of the PHAT observations {\em and}
the known distance to the stars allowing the BEAST to fit luminosities
instead of fluxes.

\subsection{Example SED Fit}

Fig.~\ref{fig_indiv_sed_fit} gives an example of the BEAST fit to a
single source in PHAT Brick 21.  This particular source was picked as
it has all 6 bands detected with positive flux and represents a common
source found in the PHAT survey (i.e., a red clump star).  The
components of the best fitting models are shown 
illustrating the strong impact of the photometric measurement bias on
the observed flux.  The 1D pPDFs show a range of behaviors.  The
$log(M)$, $log(t)$, $R(V)$, $\log(T_\mathrm{eff})$, and $\log(g)$
show single peaked pPDFs.
The $A(V)$ pPDF shows a double peaked pPDFs.  Finally, the
$f_\mathcal{A}$ and $Z$ pPDFs show sloped 
pPDFs with no clear peak.  This wide range of non-Gaussian and
multi-peaked pPDFs illustrates the complexity of this type of SED
fitting and the importance of fully mapping the pPDF.  This complexity
is one of the primary motivations for basing the BEAST on a grid
fitting technique.

\subsection{Recovered Hertzsprung-Russel Diagram}

\begin{figure}[tbp] 
\epsscale{1.35}
\plotone{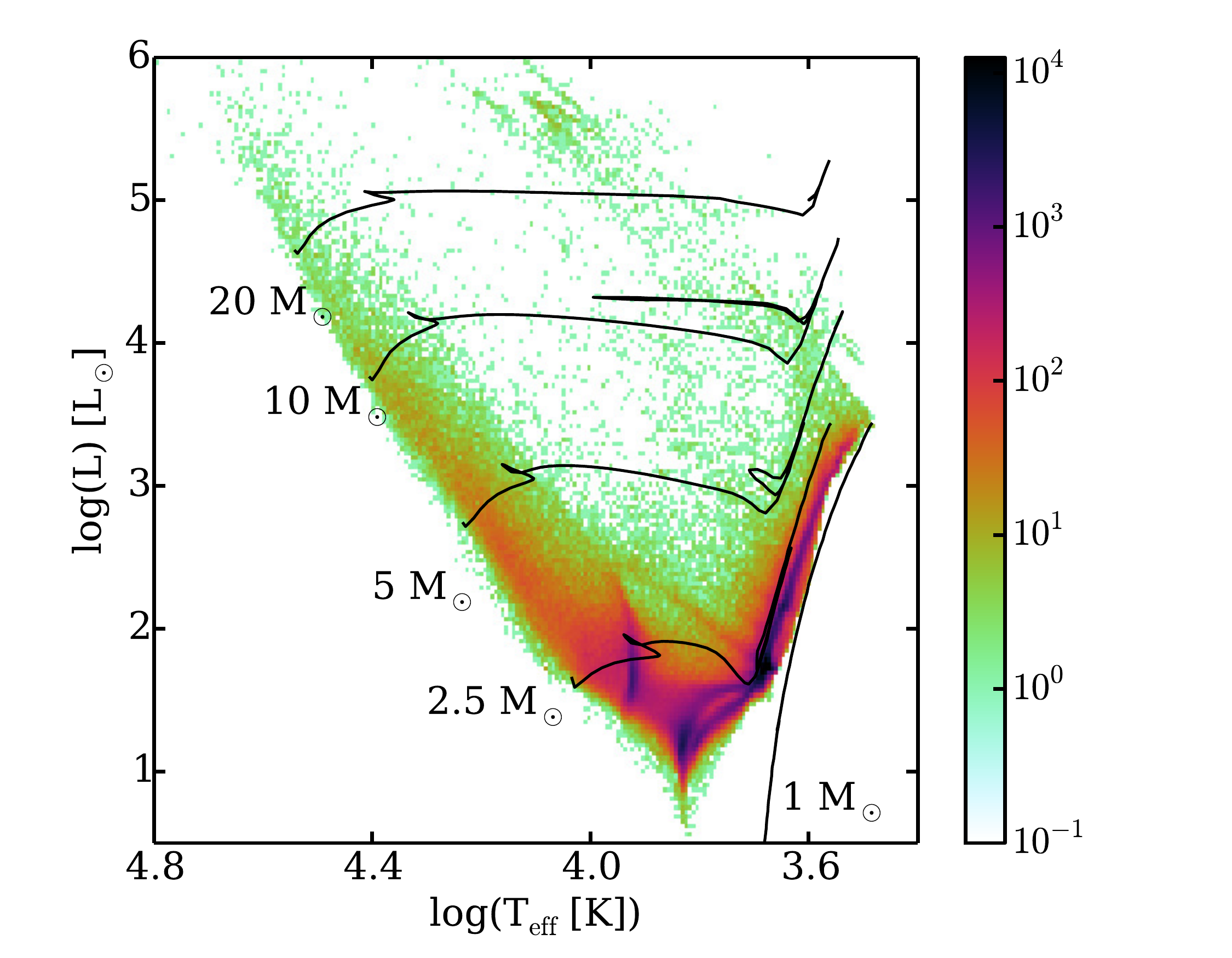} 
\caption{Using the Brick 21 results, the density of sources is shown
  in a Hertzsprung Russell diagram.  The overall structure is
  reasonable, but there are nonphysical streaks in the upper portion
  and ridges in the lower portion.  The nonphysical ridges in the faint source
  population are due to the sensitivity limits of the PHAT survey
  sources used.  We required each source to be detected in four PHAT
  bands and to have a F475W magnitude above 27.6.  In general, the
  particular set of four PHAT bands fulfilling this requirement varies
  from being UV/optical to optical/NIR and this is the expected origin
  of the two prominent ridges in the faint source population.  The
  nonphysical steaks for the brighter sources are likely due to
  variable survey depth due to gaps between the chips in WFC3 and ACS
  and/or an issue with the adopted BEAST grid being too coarse in this
  region of the HR diagram.
  \label{fig_ex_hr_diagram_density}}
\end{figure}

The Hertzsprung-Russel diagram for all the Brick~21 sources fit is given in
Fig.~\ref{fig_ex_hr_diagram_density}.  The distribution of sources is
reasonable, with sources on the RGB dominating.  The main sequence is
well populated, consistent with the large number of star forming
regions present in Brick~21.  The lower region shows the sensitivity
limit that is the result of requiring detections in at least four
bands coupled with the varying dust columns and crowding noise across
this brick.  Comparing this figure with Fig.~\ref{fig_prior_age_mass}
indicates that this region has, not surprisingly, a somewhat different
star formation history than we assumed in our priors.  This difference
provides
confirmation that the BEAST results are not dominated by the 
priors.

\subsection{Fit Uncertainties}

\begin{figure*}[tbp] 
\epsscale{1.1}
\plottwo{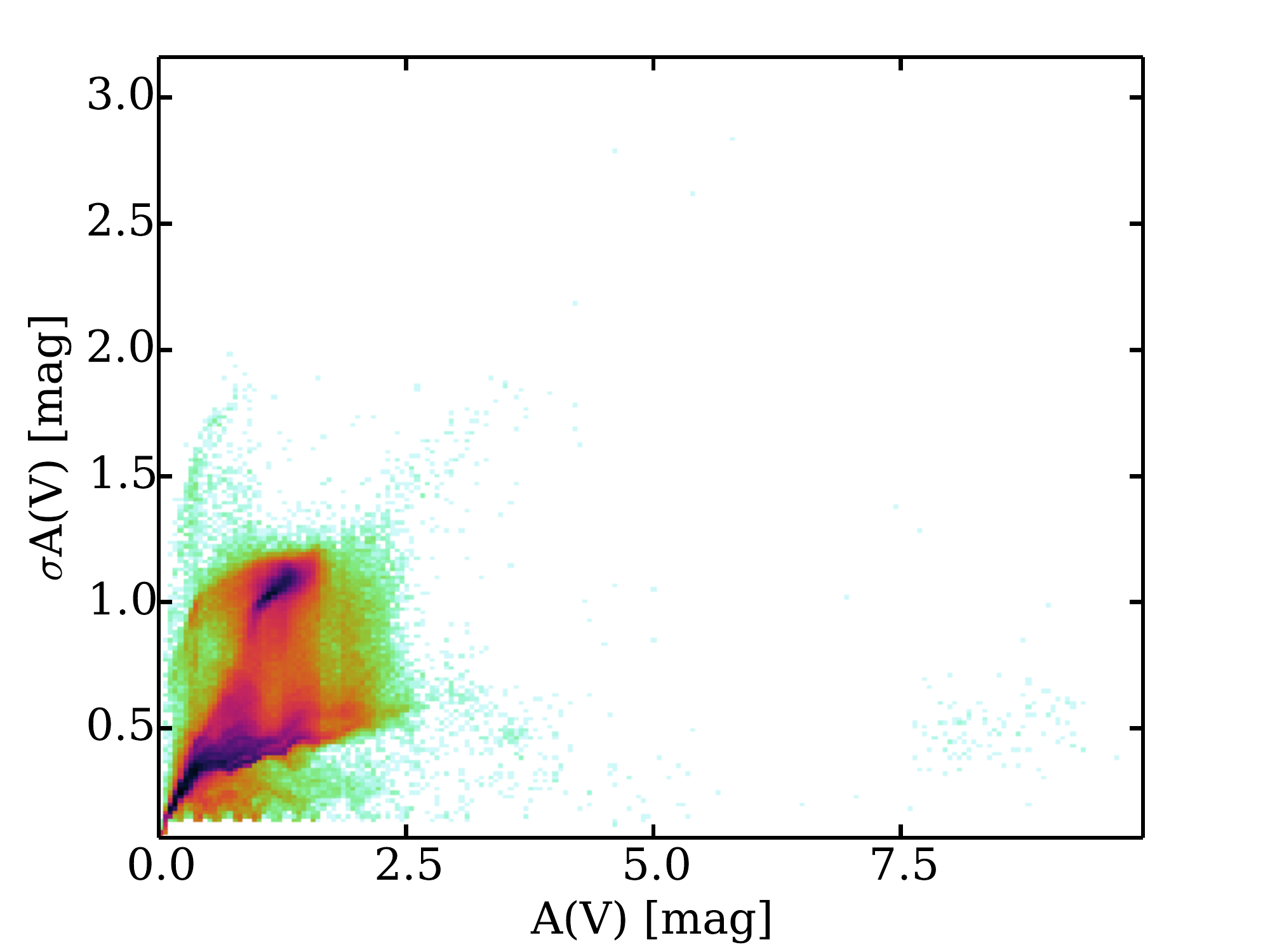}{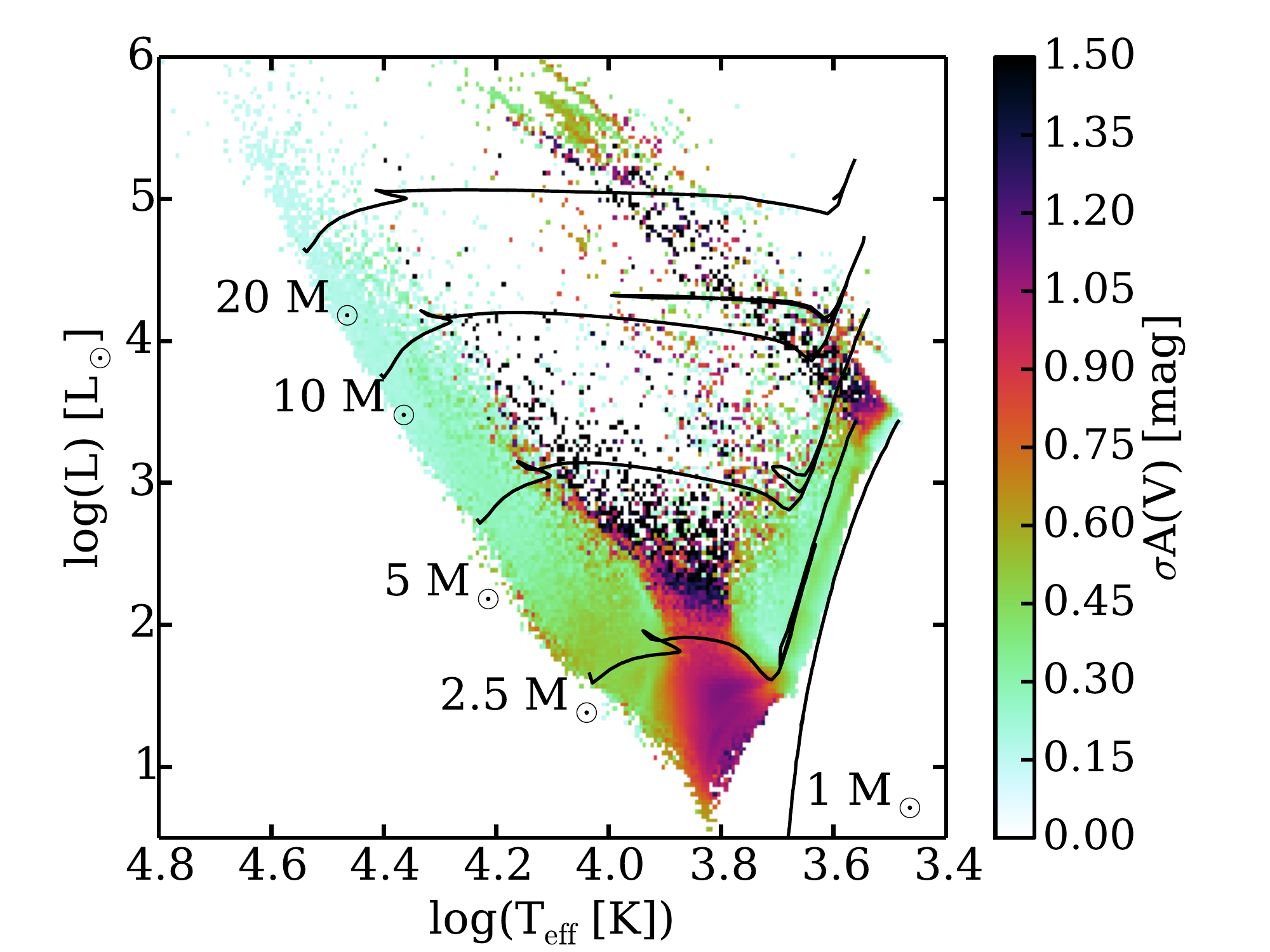} \\
\plottwo{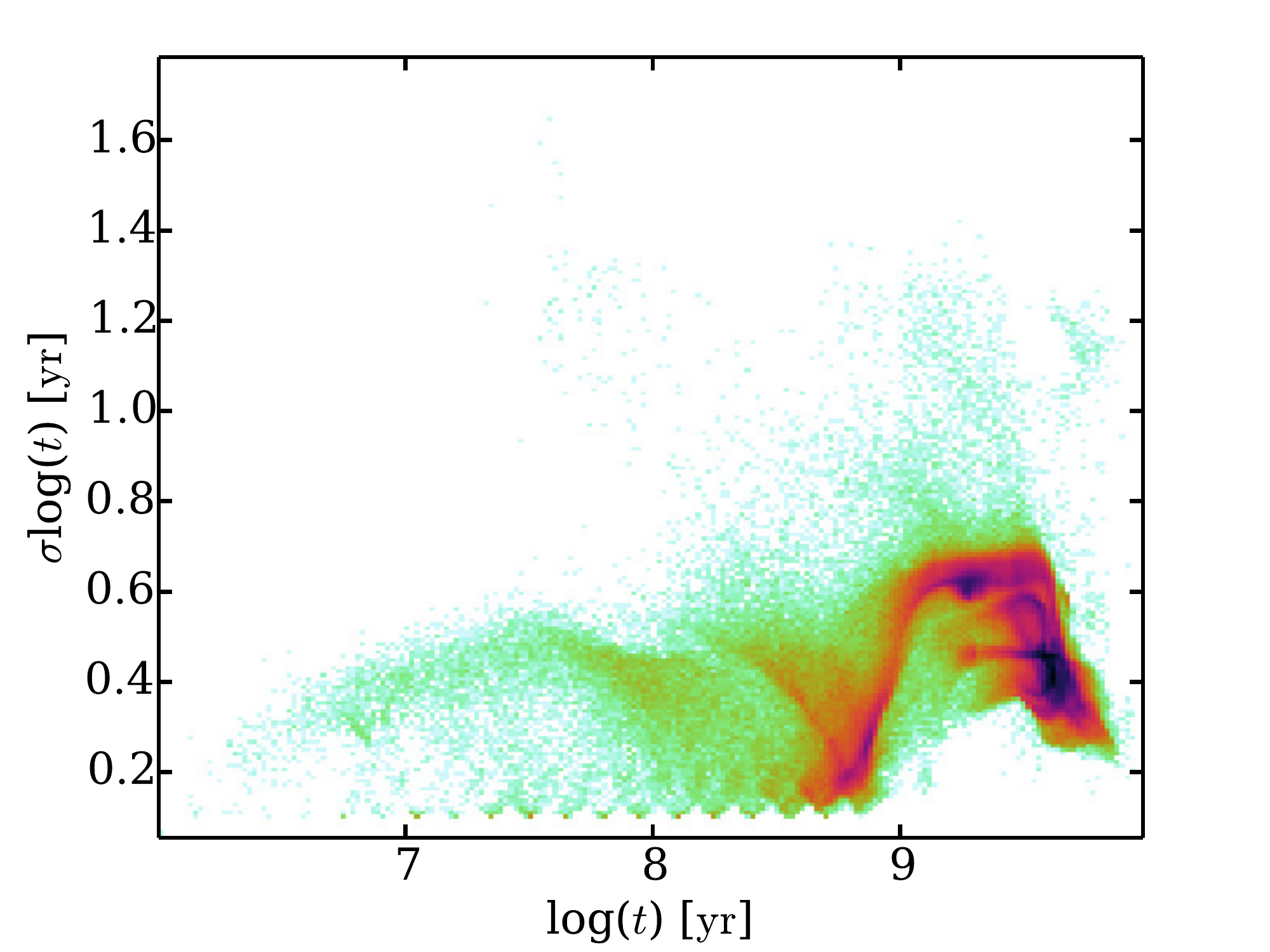}{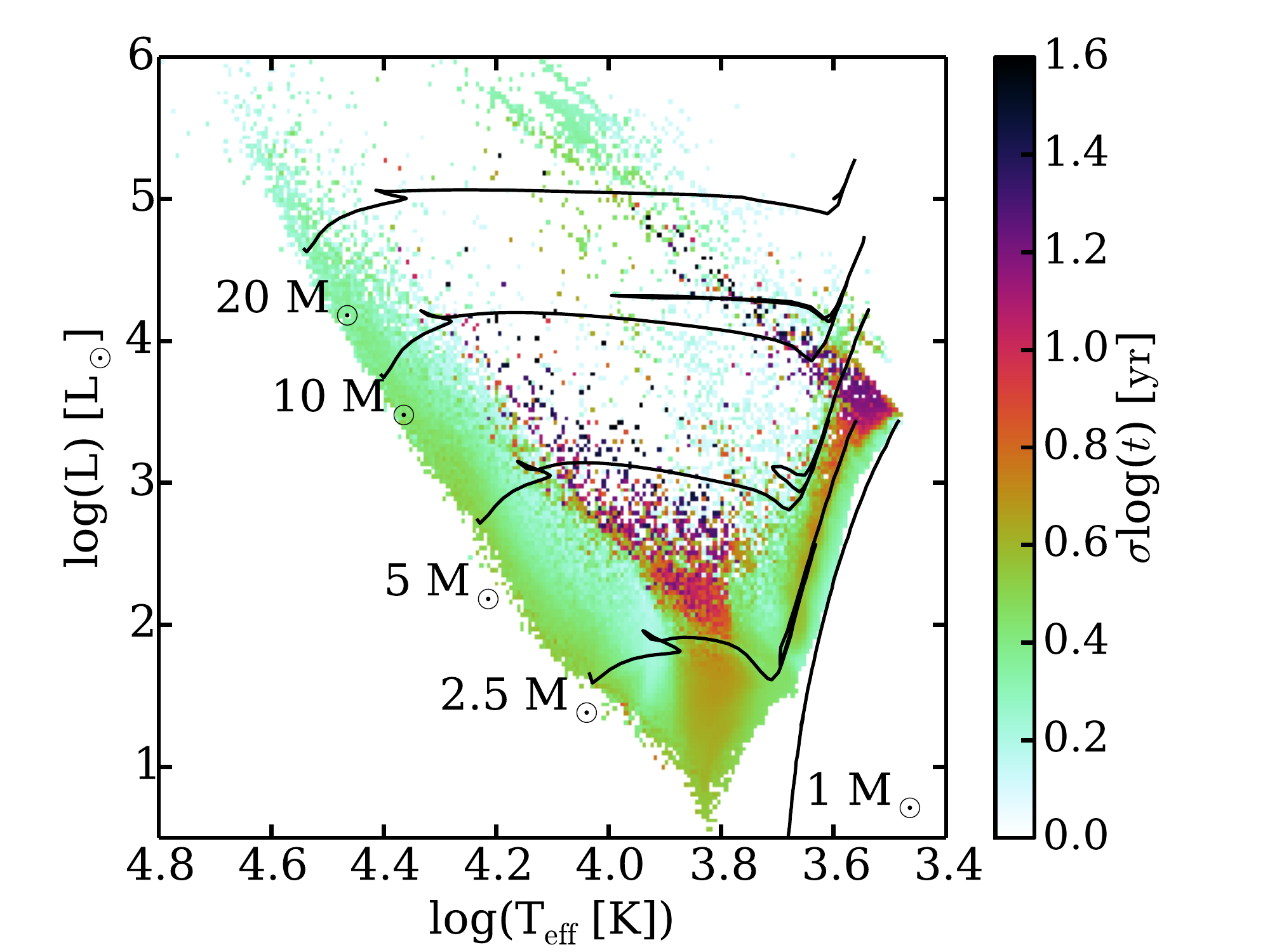} \\
\plottwo{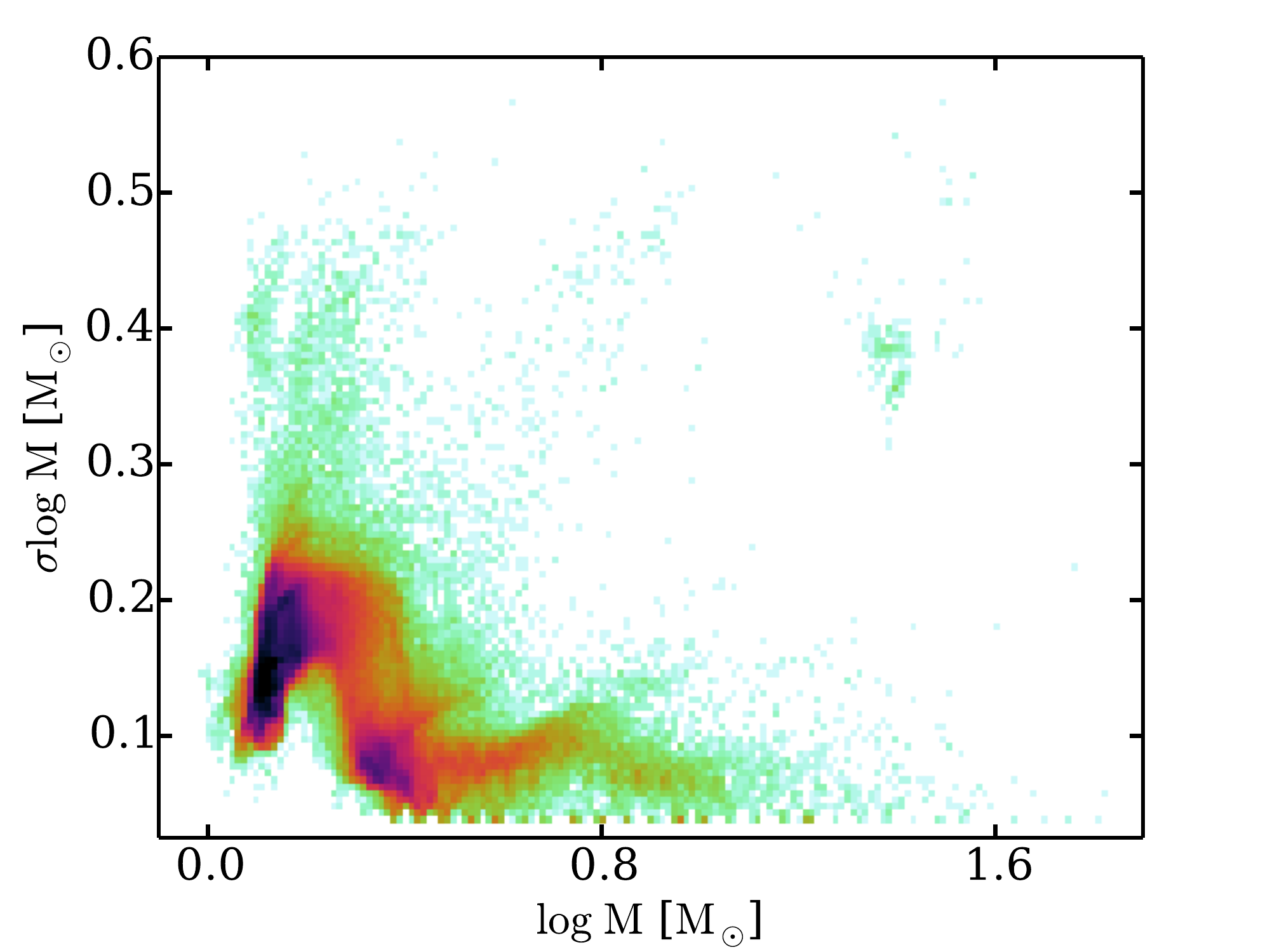}{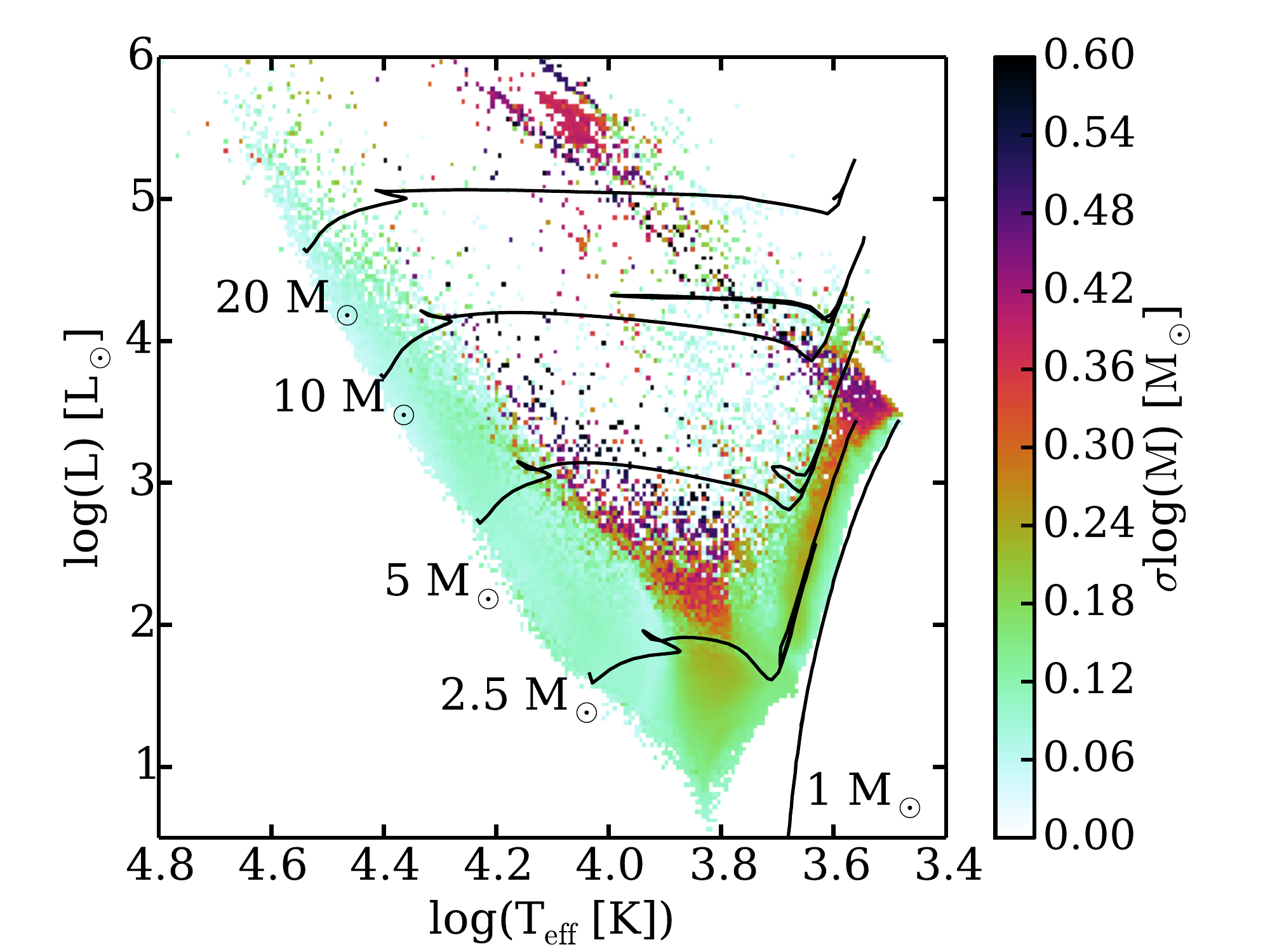} 
\caption{In the left column, the 1$\sigma$ uncertainties are plotted
  versus the expectation values for the primary fit parameters color
  coded by log density (green low, black high) from the Brick 21
  fitting.  In the right column, the average 1$\sigma$ uncertainty at
  each position in a HR diagram is shown.  The careful reader will
  notice that the right $t$ plot looks like the Loch Ness Monster.
  \label{fig_ex_fit_v_unc_primary}}
\end{figure*}

\begin{figure*}[tbp] 
\epsscale{1.1}
\plottwo{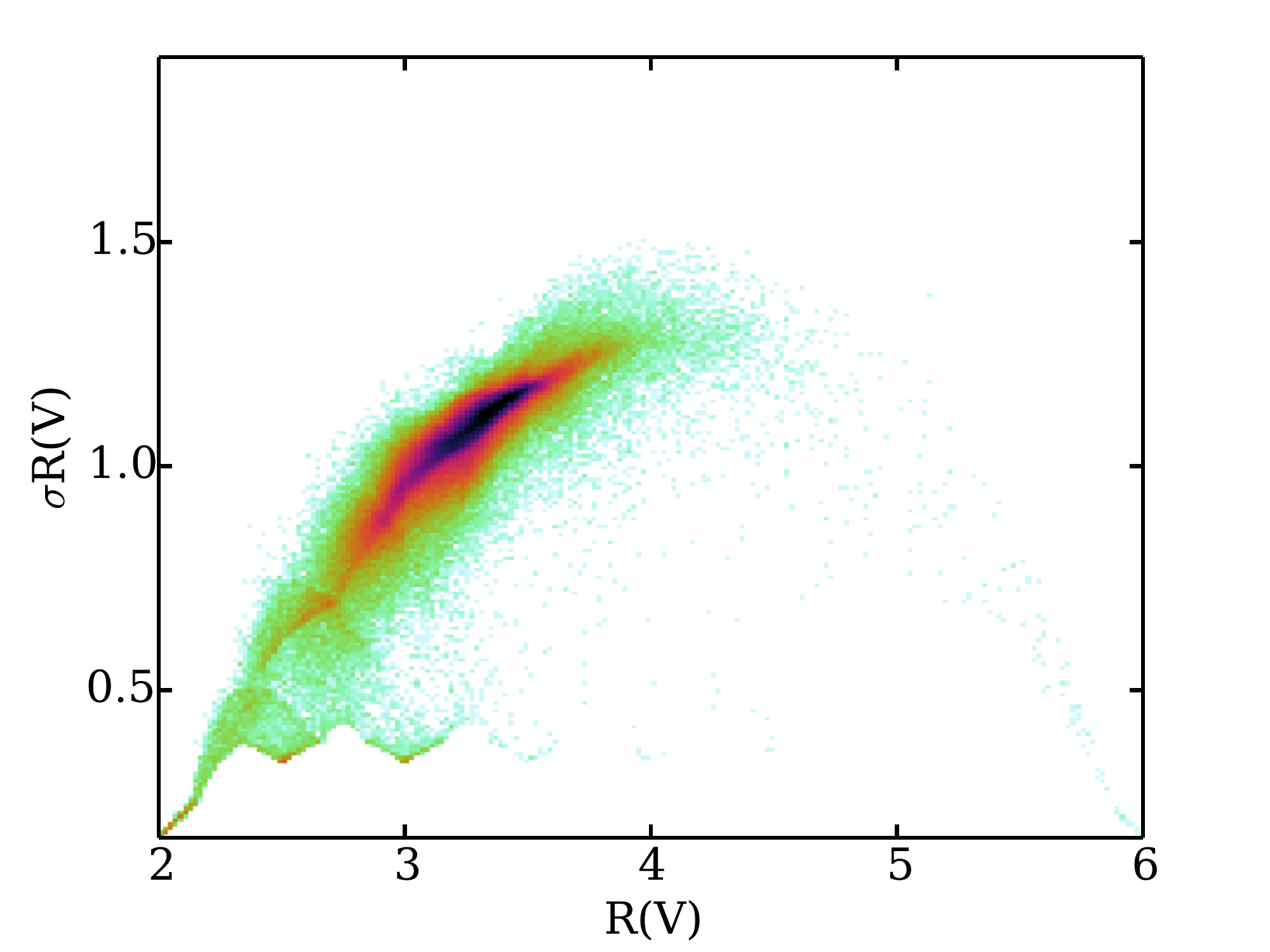}{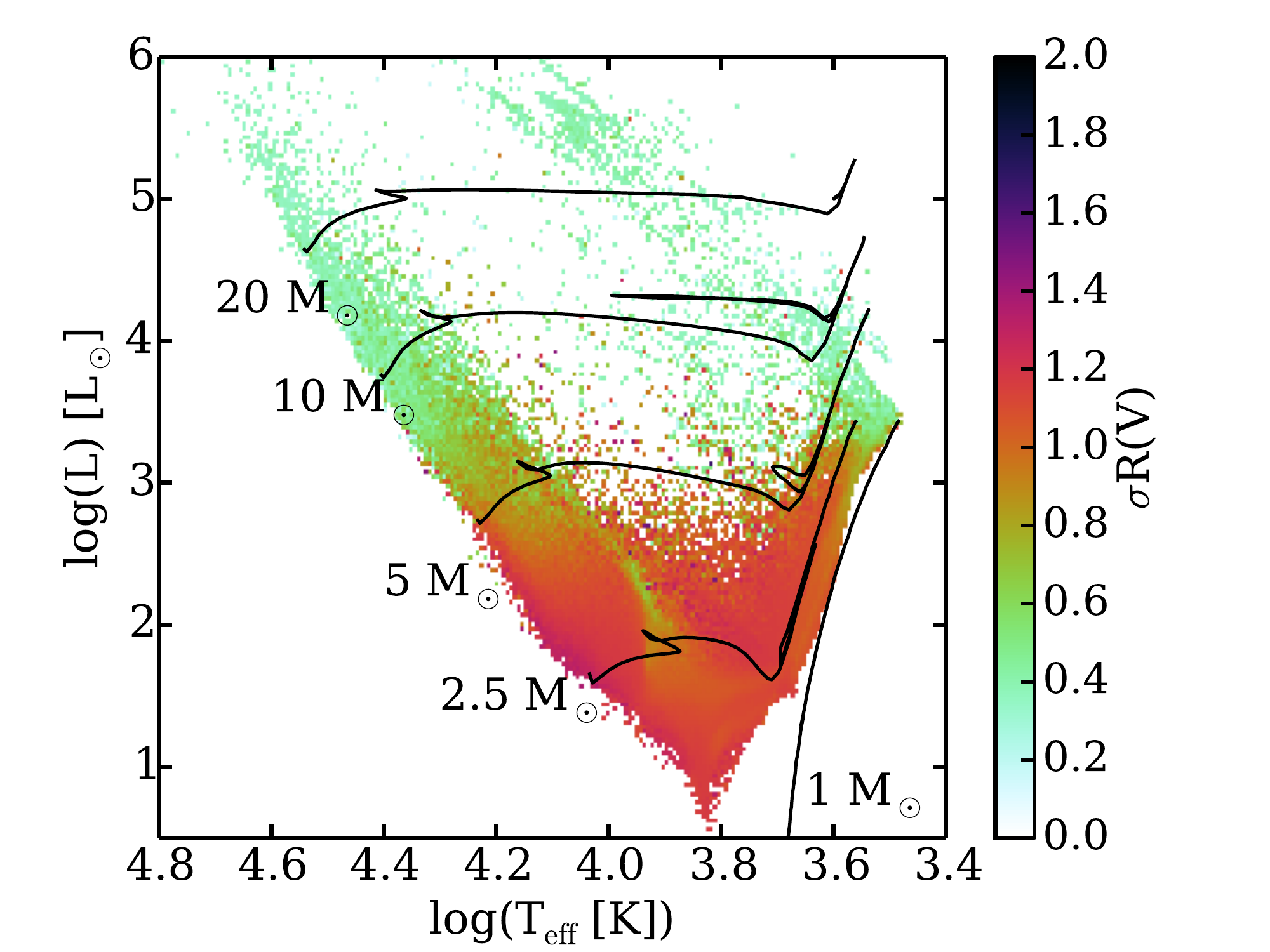} \\
\plottwo{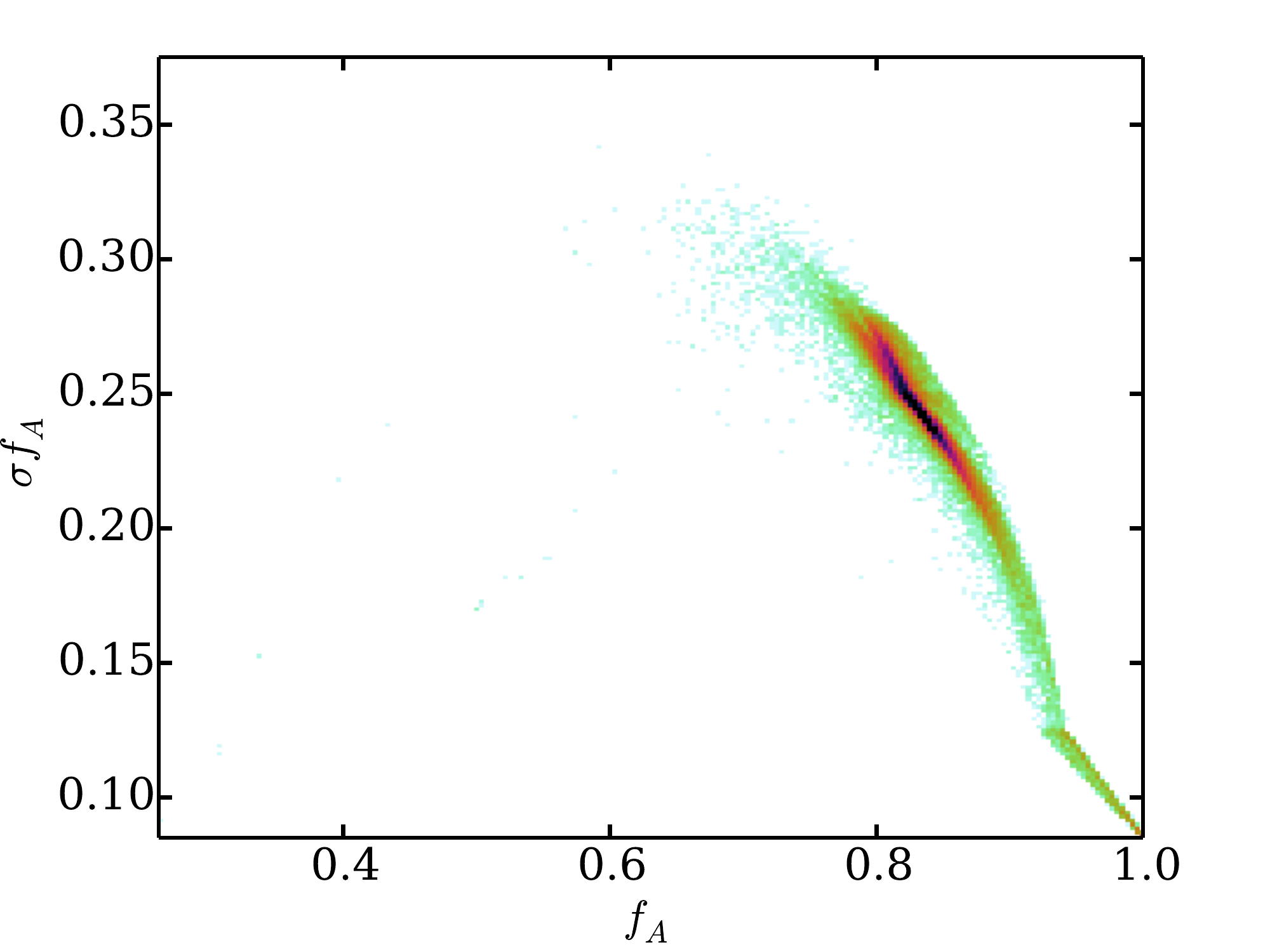}{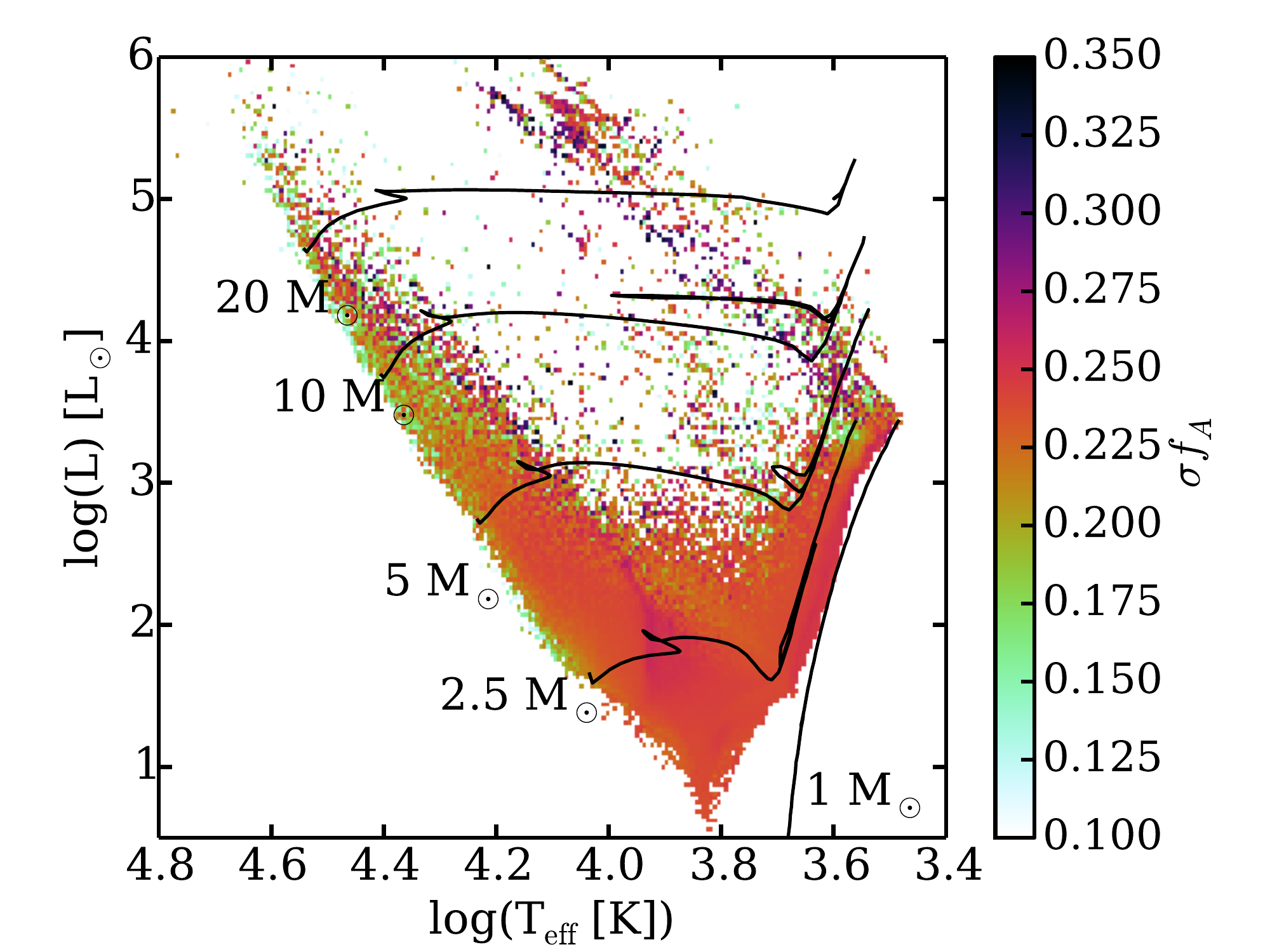} \\
\plottwo{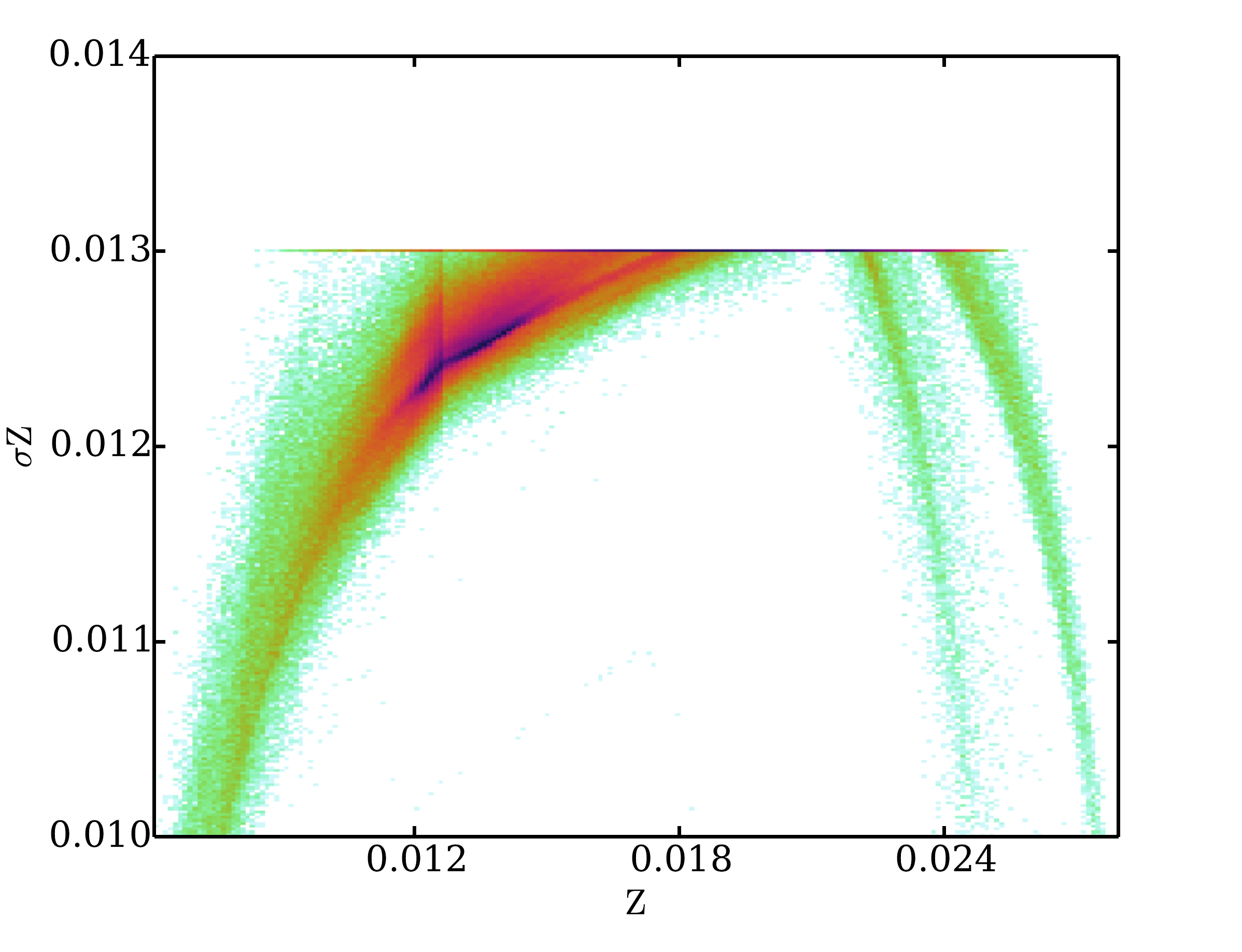}{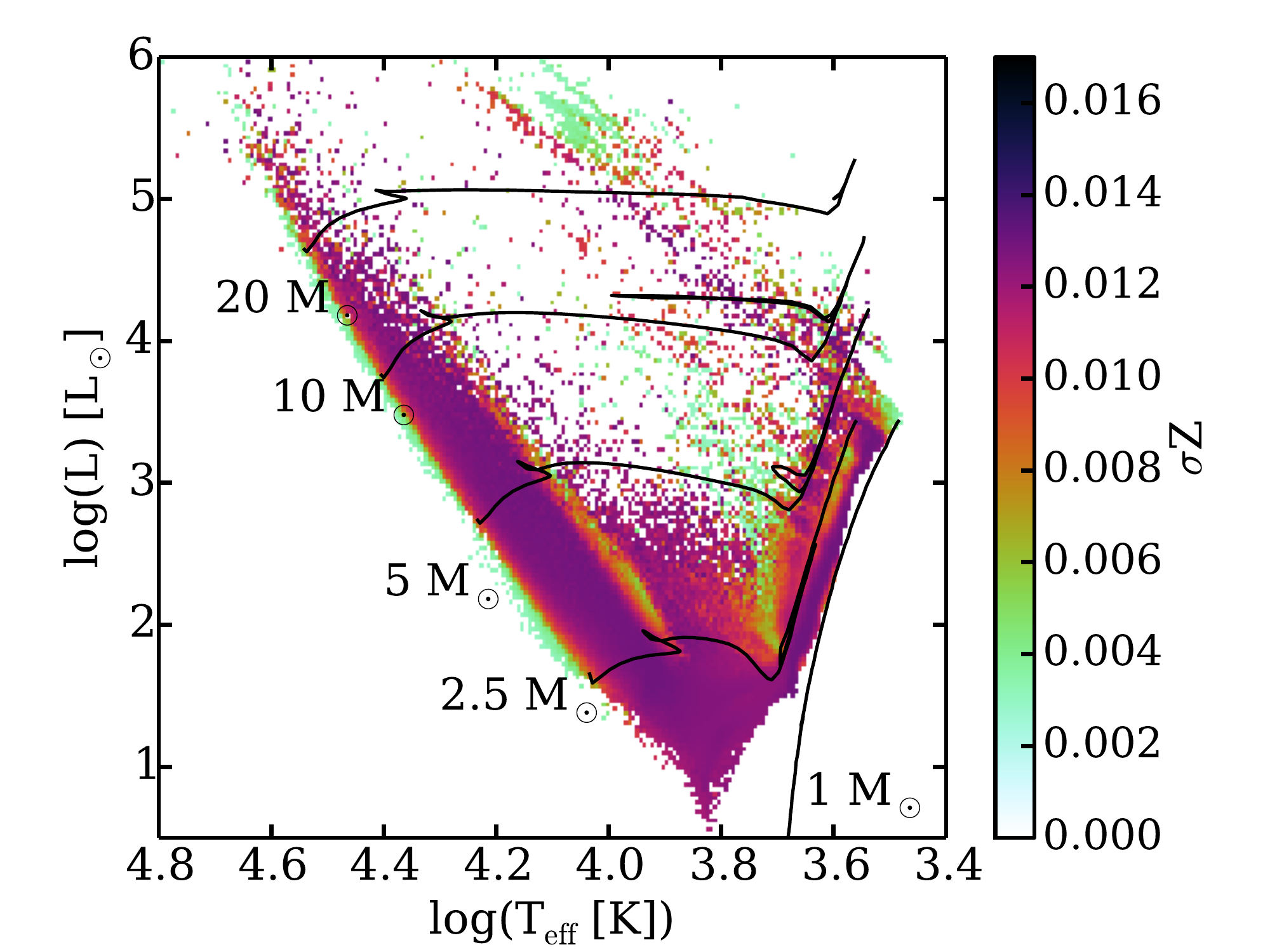} 
\caption{In the left column, the 1$\sigma$ uncertainties are plotted
  versus the expectation values for the secondary fit parameters color
  coded by log density (green low, black high) from the Brick 21
  fitting.  In the right column, the average 1$\sigma$ uncertainty at
  each position in a HR diagram is shown.  The sharp upper boundary in
  $\sigma_Z$ is the when the pPDF is completely unconstrained and is
  one half the allowed range in $Z$.
  \label{fig_ex_fit_v_unc_secondary}}
\end{figure*}

\begin{figure*}[tbp] 
\epsscale{1.1}
\plottwo{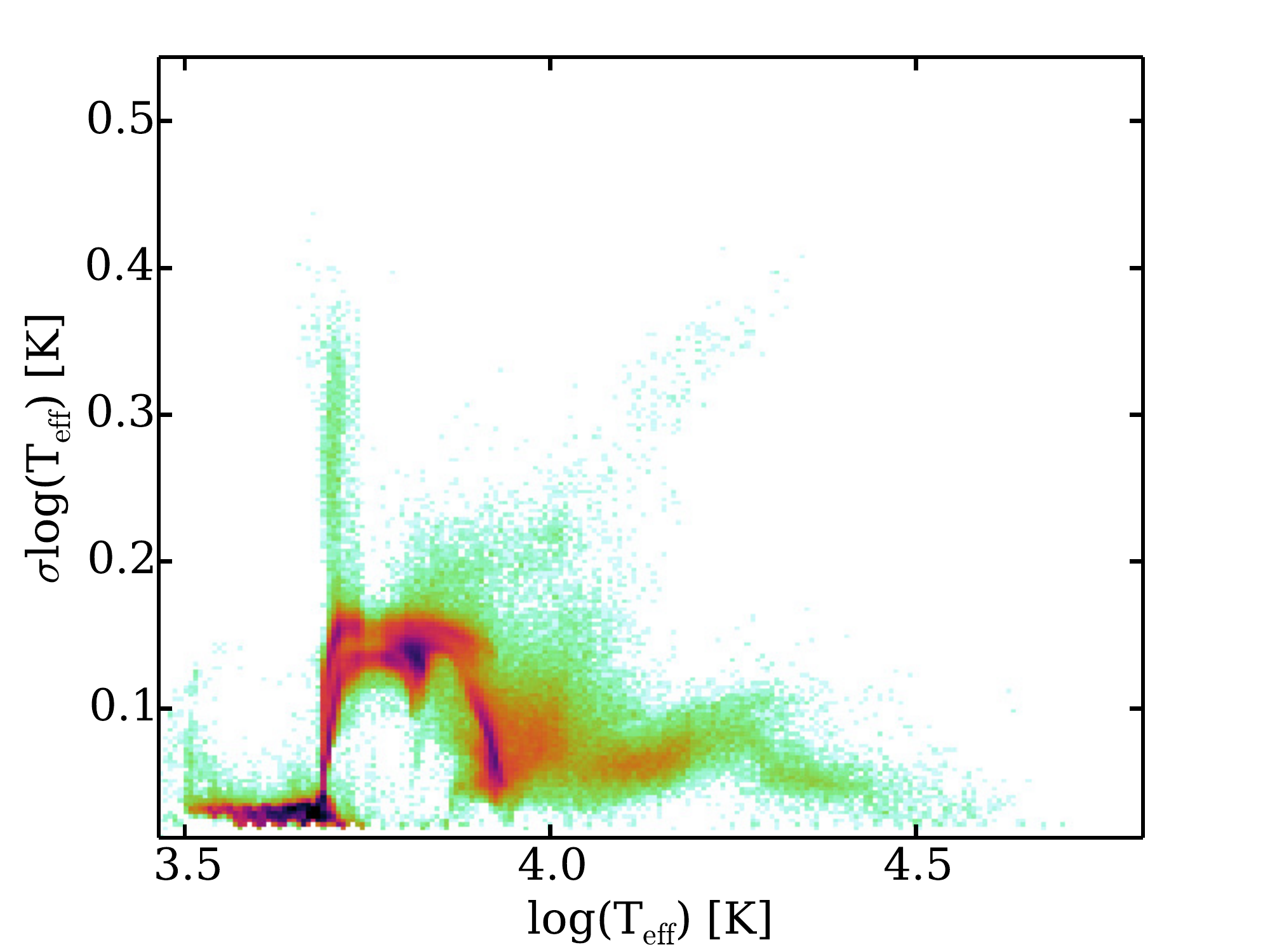}{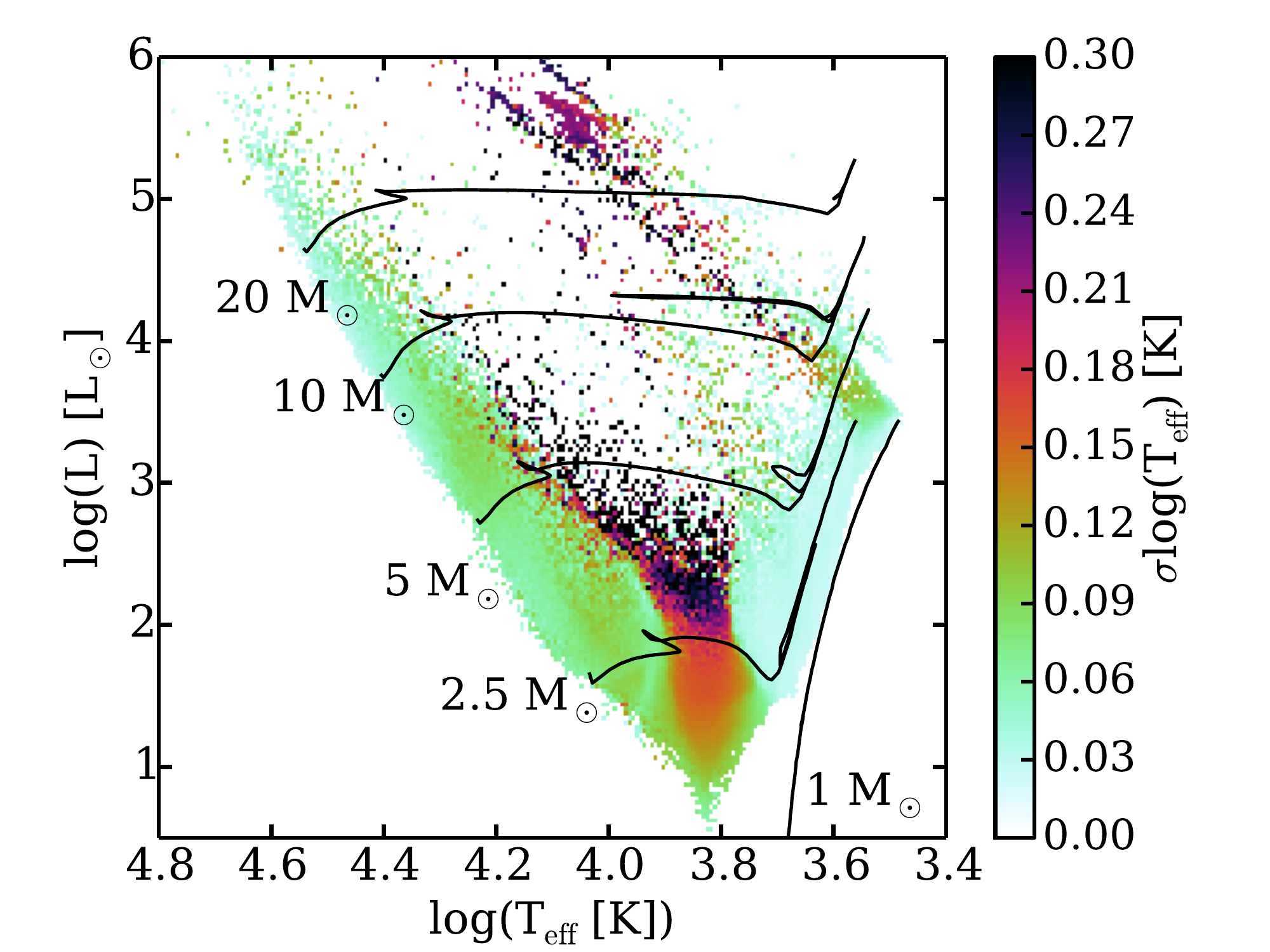} \\
\plottwo{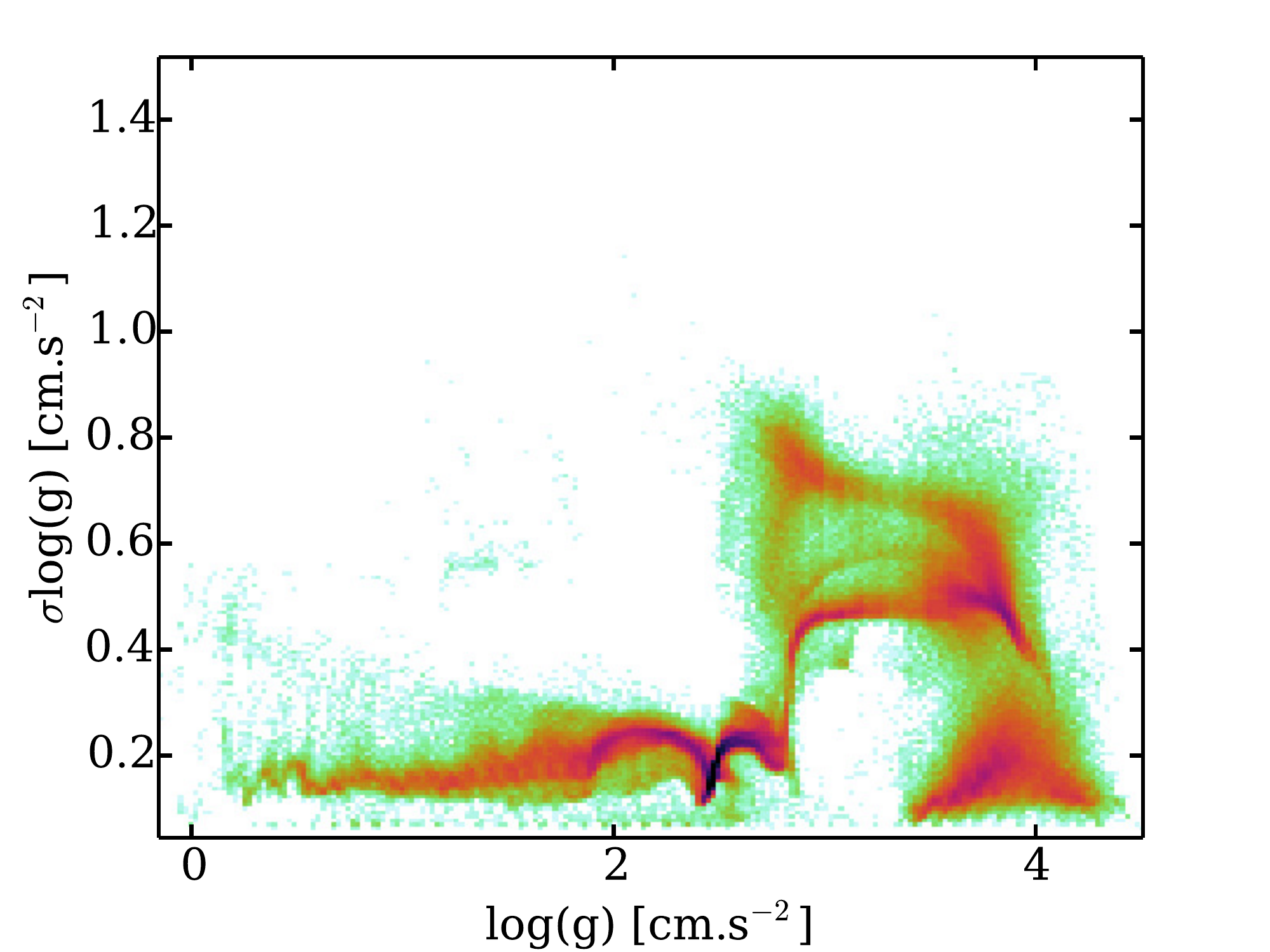}{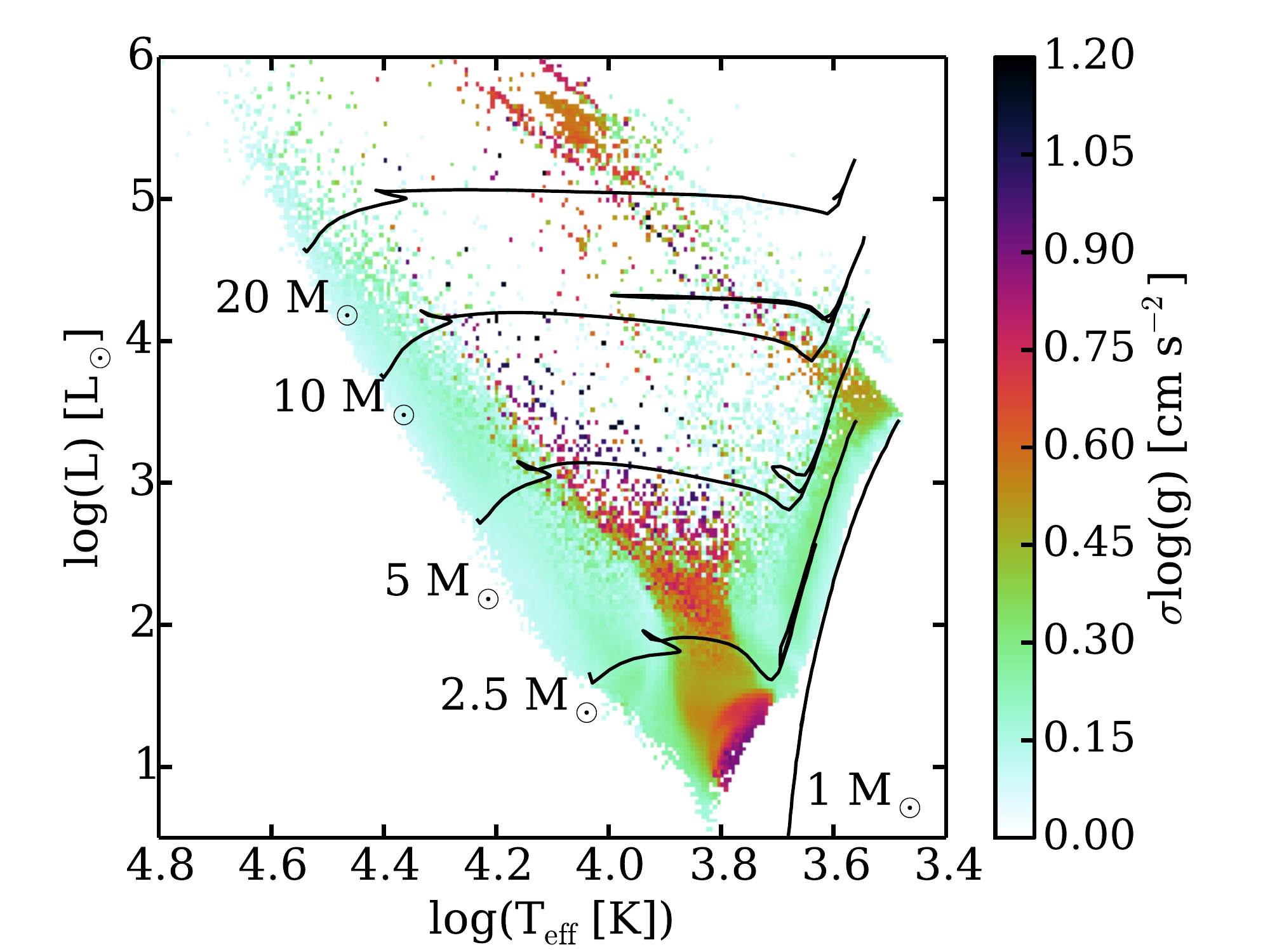} 
\caption{In the left column, the 1$\sigma$ uncertainties are plotted
  versus the expectation values for the derived fit parameters color
  coded by log density (green low, black high) from the Brick 21
  fitting.  In the right column, the average 1$\sigma$ uncertainty at
  each position in a HR diagram is shown.
  \label{fig_ex_fit_v_unc_derived}}
\end{figure*}

For all the 0.7 million sources fitted in Brick~21,
Figs.~\ref{fig_ex_fit_v_unc_primary},
\ref{fig_ex_fit_v_unc_secondary}, \& \ref{fig_ex_fit_v_unc_derived}
show the primary, secondary, and derived 1$\sigma$ fit parameter
uncertainties versus their expectation values.  In addition these
figures give the average fit parameter uncertainty color coded on HR
diagrams.  The 1$\sigma$ uncertainties are measured from the 67\%
width of the 1D pPDFs.  These figures illustrate the precision that is
possible with the BEAST for the PHAT observations of the Brick~21
region for sources detected in at least four bands with the assumed
physical and noise models.  These figures provide a detailed view
how the fit parameter precision versus fit parameter and location in
the HR diagram.  The fit parameter accuracy is discussed in
\S\ref{sec:sensitivities} 
as this can only be tested with simulations.  The precision in both
the real and simulated data for this Brick are similar.

Fig.~\ref{fig_ex_fit_v_unc_primary} clearly illustrates that the
precision in the recovery of the primary parameters is good, but
variable.  For example, the $A(V)$ plots show a bimodal precision with
concentrations at $\sim$0.4 and $\sim$1.0~mag.  The higher precision
results can be seen via the HR diagram to be the result of sources on
the main sequence and red giant branch.  The low precision results are
due to sources at the tip of the RGB as well as the faintest sources.
The results for the RGB tip sources is likely due to the lack of
Asymptotic Giant Branch stars in the BEAST model as they are missing
from the PARSEC isochrones used.  The results for the faint sources
are simply the result of low signal-to-noise measurements.  The
results for the other two primary parameters are similar in that the
RGB tip sources and faint sources are less well recovered.  But there
are differences.  The $\log(M)$ plots show that the mass is well
recovered on the main sequence, and less well recovered on the red
giant branch.  The combined low precision results for $\log(M) < 0.3$
and $\log(t) > 9$ can be traced to RGB and Red Clump stars being very
degenerate in the HR diagram for all masses between 1 and 2 $M_\odot$
at the precision of the PHAT observations.

The precision in the recovery of the secondary parameters shown in
Fig.~\ref{fig_ex_fit_v_unc_secondary} is fairly well behaved with a
fairly clear relationship between parameter and uncertainty.  The
HR diagram for $R(V)$ shows a strong gradient from bright to faint
sources, illustrating that the $R(V)$ precision is driven by the
signal-to-noise in the observations.  The $f_\mathcal{A}$ and $Z$
HR diagrams show fairly low precisions throughout
indicating the
overall difficulty in recovering these two parameters given the PHAT
bands and depth.

Tthe precision of the recovery of the derived parameters
(Fig.~\ref{fig_ex_fit_v_unc_derived}) show behaviors similar to those
seen for the primary parameters.  The main sequence and RGB stars have
better precision than the RGB tip and the fainter sources.  The
behavior for lower precision for sources with $\log(T_\mathrm{eff})$
values between 3.75 and 3.9 is a result of residual degeneracies
between $T_\mathrm{eff}$ and $A(V)$ \citep{Romaniello02}.

It is worth remembering that these precisions are based on a noise
model that does not take into account any covariance.  We have shown
in \S\ref{sec_import_cov} that once we have ASTs measured
simultaneously in all 6 bands, we will be able to improve the
precision of the BEAST fits and potentially reduce the degeneracies
between parameters.

\subsection{Dust Maps}

\begin{figure}[tbp]
\epsscale{1.4} 
\plotone{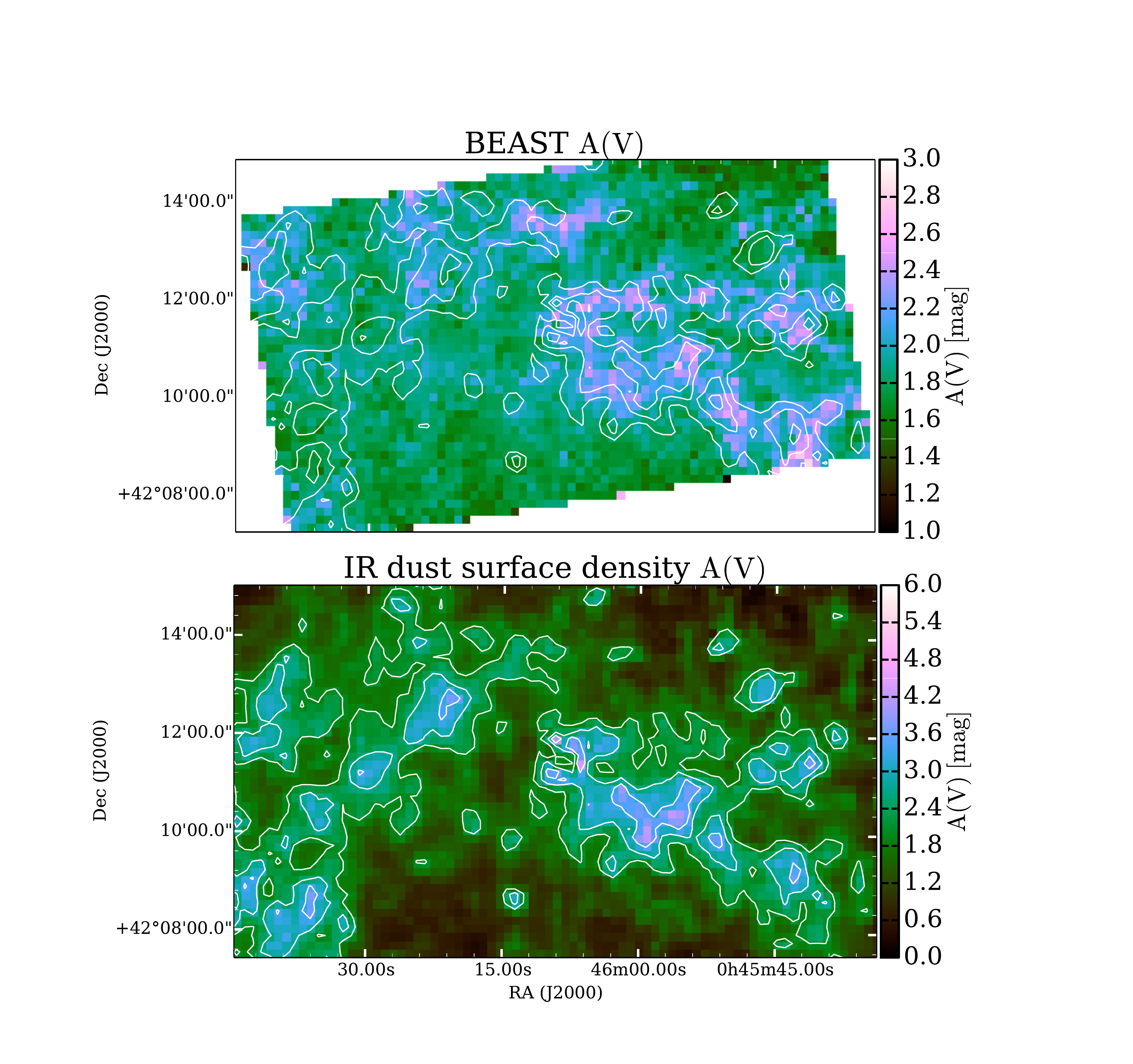}
\caption{The $A(V)$ map derived from averaging the BEAST results for
  PHAT brick 21 (top) and the IR derived dust surface density
  \citep[bottom,][]{Draine14} are shown.  The contours in both panels
  show the IR derived dust surface density.  The overall scale is 2x
  higher in the IR derived dust map as the \citet{Draine14} prediction
  of $A(V)$ is known to be systematically 2x higher due 
  to dust grain model calibration issues \citep{Lombardi14,
    Dalcanton15, Planck16}.  The two panels show very similar
  morphologies. 
  \label{fig_av_comp}}
\end{figure}

One of the specific goals of the BEAST effort is to study the dust
properties in galaxies.  A map of the dust column density can be
derived by averaging the results for individual stars in rectangular
regions.  This gives an estimate of the average dust column density in
these regions after multiplication by a factor of two as this corrects
for the fact that (on average) half of the stars will be behind the
dust and the other half in front in a disk galaxy.  This was done for
the PHAT Brick~21 sources and the resulting $A(V)$ map is given in
Fig.~\ref{fig_av_comp}.  There are coherent structures in the $A(V)$
map and similar structures are also seen in the dust surface density
map derived from fitting the dust emission infrared maps
\citep{Draine14}.  The variations between these two images provide
information on real variations in the IR dust emissivity modulo
systematics in the BEAST derived $A(V)$ map (e.g., sensitivity to the
highest columns of dust).  In future work, we will derive a more
accurate $A(V)$ map using a detailed dust and star geometry model, the
full pPDFs, and accounting for the completeness of the PHAT
observations (Arab et al., in prep.).  The comparison in this paper is
presented to show that the BEAST produces reasonable results for dust
column densities.

\section{Summary}

We have presented the BEAST, a probabilistic approach to modeling the
photometric SEDs of sources detected in large resolved star surveys.
The BEAST uses a 6 parameter model of the SED of individual stars
extinguished by dust.  The stellar portion of the model is based on
stellar evolution and atmosphere models and has the parameters of age
($t$), mass ($M$), and metallicity ($Z$).  The dust extinction portion
of the model is based on a newly developed mixture model of dust
extinction and has the parameters of dust column ($A(V)$), average
dust grain size ($R(V)$), and mixture coefficient ($f_A$).  This
mixture model has the feature of encompassing all observed extinction
curves in the Milky Way, Large Magellanic Cloud, and Small Magellanic
Cloud.  

The BEAST fitting technique allows for bias and covariance between
bands in the observed SEDs using an multi-variate Normal/Gaussian
distribution.  The noise model used includes terms for the photon,
crowding, and absolute calibration noise. The importance of correctly
integrating the model SEDs using the full photometric band response
functions was illustrated.  The impact on the fit parameter precision
and accuracy of correctly accounting for the covariance between
observed bands in the noise model was shown.  Overall, the strength of
the BEAST over most existing SED fitting codes is a careful treatment
of biases and correlations in the uncertainties between bands and the
new dust extinction mixture model.

We illustrated the application of the BEAST to real data using the
PHAT survey of M31.  The assumed priors are based on independent
knowledge of stellar and dust physics and were presented in the
context of fitting using a grid.  The derivation of the noise model
from ASTs and the details of the HST absolute flux calibration were
described.  Using sensitivity tests, we illustrated that overall the
BEAST does a good job recovering the input fit parameters.  Finally,
we presented the results from fitting all the sources detected in 4
bands in Brick~21 of the PHAT survey.  These results yield a
reasonable HR diagram, understandable variations in fit parameter
precision, and a spatial map
of the average dust column that correlates well with existing infrared
derived dust surface densities.

While the BEAST has been motivated and will be used for the PHAT
survey to study the stellar and dust content of M31, it can be used to
fit the data from any resolved star extragalactic survey.

\acknowledgements

We thank the referee for an insightful comments that motivated
  us to significantly improve the content and presentation of this paper.
Support for program \# 12055 was provided by NASA through a grant from
the Space Telescope Science Institute, which is operated by the
Association of Universities for Research in Astronomy, Inc., under
NASA contract NAS 5-26555.  Support for DRW is provided by NASA
through Hubble Fellowship grants HST-HF-51331.01 awarded by the Space
Telescope Science Institute.  This work used the Extreme Science and
Engineering Discovery Environment (XSEDE), which is supported by
National Science Foundation grant number ACI-1053575.  The authors
acknowledge the Texas Advanced Computing Center (TACC) at The
University of Texas at Austin for providing High Performance Computing
(HPC) resources that have contributed to the research 
results reported within this paper. URL: http://www.tacc.utexas.edu.

\end{document}